\documentclass{jfm}
\usepackage{graphicx}
\usepackage{epstopdf, epsfig}
\usepackage{hyperref}
\usepackage{amsmath}
\usepackage{amssymb}
\usepackage{placeins}
\usepackage{color,soul}
\usepackage{enumerate}
\usepackage{bm}
\usepackage{natbib}

\newcommand{\hla}[2][white]{{\sethlcolor{#1}\hl{#2}}}
\newcommand{\hlb}[2][white]{{\sethlcolor{#1}\hl{#2}}}
\newcommand{\hlc}[2][white]{{\sethlcolor{#1}\hl{#2}}}

\title{Uncertainty in Elastic Turbulence}

\author{J. R. C. King\aff{1}
  \corresp{\email{jack.king@manchester.ac.uk}},
  R. J. Poole\aff{2},
  C. P. Fonte\aff{3},
 \and S. J. Lind\aff{4}}

\affiliation{\aff{1}Department of Mechanical and Aerospace Engineering, The University of Manchester, Manchester, UK
\aff{2}School of Engineering, University of Liverpool, Liverpool, UK
\aff{3}Department of Chemical Engineering, The University of Manchester, Manchester, UK
\aff{4}School of Engineering, Cardiff University, Cardiff, UK}

\begin{document}

\maketitle
\begin{abstract}

% Background
Elastic turbulence can lead to increased flow resistance, mixing and heat transfer. Its control -- either suppression or promotion --  has significant potential, and there is a concerted ongoing effort by the community to improve our understanding. Here we explore the dynamics of uncertainty in elastic turbulence, inspired by an approach recently applied to inertial turbulence in Ge et al. (2023) \textit{J. Fluid Mech.} 977:A17. We derive equations for the evolution of uncertainty measures, yielding insight on uncertainty growth mechanisms. Through numerical experiments, we identify four regimes of uncertainty evolution, characterised by I) rapid transfer to large scales, with large scale growth rates of $\tau^{6}$ (where $\tau$ represents time), II) a dissipative reduction of uncertainty, III) exponential growth at all scales, and IV) saturation. These regimes are governed by the interplay between advective and polymeric contributions (which tend to \hla{increase} uncertainty), viscous, relaxation and dissipation effects (which reduce uncertainty), and inertial contributions. In elastic turbulence, reducing Reynolds number increases uncertainty at short times, but does not significantly influence the growth of uncertainty at later times. At late times, the growth of uncertainty increases with Weissenberg number, with decreasing polymeric diffusivity, and with the logarithm of the maximum length scale, as large flow features adjust the balance of advective and relaxation effects. These findings provide insight into the dynamics of elastic turbulence, offering a new approach for the analysis of viscoelastic flow instabilities.

\end{abstract}

\maketitle
%end front matter

\section{Introduction}\label{sec:intro}

%% Rough background to elastic turbulence
Elastic turbulence~\citep{groisman_2000}, a chaotic flow state observed in polymer solutions even in the limit of vanishing inertia, has implications across a range of application areas. It has been shown that elastic turbulence can enable emulsification~\citep{poole_2012}; that it can promote heat transfer~\citep{traore_2015}; that it is involved in melt fracture~\citep{morozov_2007}; and that in mass-transfer limited regimes it can increase apparent reaction rates in porous reactors~\citep{browne_2024} to provide just a few examples. Since the first documentation of elastic turbulence in $2000$ by~\cite{groisman_2000}, there has been a concerted effort amongst the community to understand the phenomenon, and we refer the reader to~\cite{steinberg_2021}, ~\cite{datta_2022a} and~\cite{sasmal_2025} for recent and comprehensive reviews. We note the related phenomena of elasto-inertial turbulence, first described by~\cite{samanta_2013}; a chaotic state seen across a large range of Reynolds numbers, driven by the polymer dynamics, where both elastic and inertial effects are important. In this work we only consider two-dimensional low Reynolds number flows in the elastic turbulence regime, although the theory we develop is applicable to viscoelastic flows in general. For a review of elasto-inertial turbulence, we refer the reader to~\cite{dubief_2023}.

%% Ge 2023
The work we present here was inspired by the recent work of~\cite{ge_2023} on uncertainty in inertial (Newtonian) turbulence. In~\cite{ge_2023}, a remarkable depth of insight was provided through a relatively simple and intuitive approach: take two realisations of the Navier-Stokes equations, subtract them, and get evolution equations for the difference. The uncertainty was defined as the spatial average of the kinetic energy of the velocity difference $\Delta{u}_{i}$:
\begin{equation}\left\langle{E}_{\Delta}\right\rangle=\frac{1}{V}\displaystyle\int\frac{1}{2}\Delta{u}_{i}\Delta{u}_{i}dV,\label{eq1}\end{equation}
for which an evolution equation was derived. Simply by inspecting the form of the evolution equation for $\left\langle{E}_{\Delta}\right\rangle$, and performing numerical experiments in which one realisation is perturbed and the evolution of the difference is tracked, they obtained fundamental quantitative insight into the chaotic nature of turbulence: identifying a similarity regime in which the production and dissipation rates of uncertainty grow together and the uncertainty spectrum is self-similar, and finding that in the absence of an external input of uncertainty, uncertainty can only be created via compression events in the Newtonian case. For chaotic flows, $\left\langle{E}_{\Delta}\right\rangle$ is expected to grow exponentially, according to 
\begin{equation}\frac{d\left\langle{E}_{\Delta}\right\rangle}{dt}=\lambda\left\langle{E}_{\Delta}\right\rangle,\end{equation}
in which $\lambda$ is twice the maximal Lyapunov exponent of the system. \hlc{Although \mbox{\cite{ge_2023}} provided our inspiration, the concept of $\left\langle{E}_{\Delta}\right\rangle$ in studies of chaos and predictability in Newtonian turbulence predates this (e.g. \mbox{\cite{boffetta_2017}}; \mbox{\cite{berera_2018}}) dating back to the work of~\mbox{\cite{leith_1972}}. To our knowledge, this concept has not previously been applied to viscoelastic flows}. We are interested in chaotic flows of polymer solutions. Here, the system cannot be completely defined by the velocity field; a conformation tensor $c_{ij}$, which provides a macroscopic measure of the molecular deformation of polymers, is also required, and the picture is more complex.

% Some recent developments
In recent years there has been a concerted effort by the community to understand and explore the fundamental dynamics of elastic (and elasto-inertial) turbulence. Much effort has focused on the onset of instabilities, aiming to address the questions of when, where and how do viscoelastic flows transition to these chaotic states. This has led to many important discoveries, including the existence of purely elastic instabilities in straight channels~\citep{pan_2013}, the first exact travelling wave solutions~\citep{boffetta_2005,page_2020,morozov_2022}, identification of a continuous pathway between elastic and elasto-inertial turbulence~\citep{khalid_2021}, and the recently discovered polymer diffusive instability (~\cite{beneitez_2023,lewy_2024,lewy_2024a,couchman_2024}) present in wall-bounded viscoelastic flows.

%% Review a little numerical work.
Early numerical studies of elastic turbulence took two-dimensional Kolmogorov flow (a flow with a uni-directional sinusoidal forcing in a doubly-periodic domain) as the setting~\citep{berti_2008,berti_2010}, and this setting is still used in more recent studies (e.g.~\cite{garg_2021}). Even in this simple configuration, the main features observed experimentally are present: a transition to a chaotic unsteady flow state above a critical Weissenberg number; a distinct increase in power input required to sustain the flow; and an energy spectra with a power law decay having a slope steeper than $k^{-3}$, where $k$ is the wavenumber. More recently, a similar numerical setting -- doubly periodic with a cellular forcing -- has been used by~\cite{plan_2017} to estimate the Lyapunov dimension of elastic turbulence, and~\cite{gupta_2019} to study the influence of polymeric diffusivity on the dynamics of elastic turbulence.
%% Is it 2D or 3D?

%% Some of the challenges and sources of uncertainty.
Uncertainty is inherent in both real flows of polymer solutions, and numerical simulations. The conformation tensor $c_{ij}$ used to describe the macroscopic behaviour of polymers is a statistical average of microscopic polymer deformations. Although if left to rest polymers will revert to an undeformed configuration ($c_{ij}\to\delta_{ij}$), in real flows of polymers, local temperature, concentration and composition fluctuations will influence viscosities and relaxation times, and potentially create local internal stresses. Consequently there is likely to be inherent uncertainty in the precise initial conditions in any real experiment. Can we ever be certain of the \emph{exact} internal molecular configuration of polymers at the start of an experiment? On average, perhaps, but not in the microscopic detail. In inertial (Newtonian) turbulence, thermal fluctuations have recently been shown to influence the energy cascade at the Kolmogorov length-scale in incompressible settings~\citep{bell_2022}, and in compressible settings, this influence extends to larger length-scales, whilst intermittency is inhibited by thermal fluctuations across the entire dissipation range~\citep{srivastava_2025}. Such findings may have implications for under-resolution approaches such as Large Eddy Simulations (LES), where closure models are used to represent the dynamics at the smallest scales. Whether such effects are significant in elastic turbulence is unknown.

For numerical simulations, a very high resolution is required to fully resolve elastic turbulence: the pseudo-spectral simulations of~\cite{berti_2008,berti_2010} were conducted with $512^{2}$ modes, the simulations by~\cite{plan_2017,gupta_2019} used a high-order compact finite difference scheme on a grid with $1024^{2}$ elements, and recent studies by~\cite{lellep_2024} use a pseudo-spectral code with $256\times256\times1024$ modes (the higher-resolution in the wall normal direction), to explore the stability of highly elastic planar channel flows. Recently,~\cite{yerasi_2024} investigated the effect of different numerical schemes, and of polymeric diffusivity, on the large scale dynamics of numerical simulations, finding that even stable numerical simulations displaying the expected chaotic fluctuations can contain numerical errors which distort the large scale flow dynamics.

%% SOme of the questions we might be interested in.
With such uncertainty inherent in viscoelastic flows, one might logically ask: how do\hla{ aleatoric} uncertaint\hla{ies propagate} in such settings? How will a slight under-resolution in a simulation impact the observed flow dynamics? How sensitive is an experiment to the initial conditions, or uncertainty in conditions such as flow rate, temperature, or geometry? How do effects such as thermal fluctuations or concentration fluctuations influence predictability and repeatability? These questions are also important in the development of closure models (analogous to those used for inertial turbulence): if we devise some statistical closure for the small scales, allowing the elastic turbulence equivalent of Large Eddy Simulation, how does uncertainty inherent in the small scales \hla{affect} larger scales, and how is the long time evolution of a flow affected?

%% Chaotic measures of turbulence
Whilst turbulence is generally analysed via spectral approaches, such questions of uncertainty are more closely related with ideas around chaotic measures~\citep{ho_2024}, such as maximum Lyapunov exponents and attractor dimension. Estimates of how the Lyapunov exponent of inertial turbulence scales date back nearly $50$ years, to the work of~\cite{ruelle_1979,deissler_1986}, although a consensus on the scaling of the Lyapunov exponent with Reynolds number still eludes the community~\citep{ge_2023}.  In the field of elastic turbulence,~\cite{plan_2017} estimated the attractor dimension based on two-dimensional numerical experiments, proposing the attractor dimension to scale with $Wi^{\alpha}$, with $\alpha\approx0.7$. Beyond this effort, we are not aware of any work analysing elastic turbulence via such measures.

%% Our contribution
Here we apply the concepts developed in~\cite{ge_2023} to viscoelastic flows. In \S~\ref{ta} we derive evolution equations for the difference between two realisations, and from these we obtain equations governing the evolution of uncertainty in the flow and the polymer deformation. These equations provide insight into the mechanisms of uncertainty generation in viscoelastic flows in general. Focusing on the elastic turbulence regime, in \S~\ref{ne} we present a set of numerical experiments, from which we identify different regimes of uncertainty evolution, and explore how these are influenced by inertial, elastic, and length scale effects. In~\S~\ref{conc} we draw conclusions and provide a discussion on future directions.

\section{Theoretical analysis}\label{ta}

We consider the incompressible isothermal flow of a polymer solution in a periodic domain. In dimensionless form, conservation equations for mass and momentum are
\begin{equation}\frac{\partial{u}_{i}^{\left(m\right)}}{\partial{x}_{i}}=0\label{eq:mass}\end{equation}
\begin{equation}\frac{\partial{u}_{i}^{\left(m\right)}}{\partial{t}}+u_{j}^{\left(m\right)}\frac{\partial{u}_{i}^{\left(m\right)}}{\partial{x}_{j}}=-\frac{\partial{p}^{\left(m\right)}}{\partial{x_{i}}}+\frac{\beta}{Re}\frac{\partial^{2}u_{i}^{\left(m\right)}}{\partial{x}_{j}\partial{x}_{j}}+\frac{\left(1-\beta\right)}{ReWi}\frac{\partial{c}_{ji}^{\left(m\right)}}{\partial{x}_{j}}+f_{i}^{\left(m\right)},\label{eq:mom}\end{equation}
where the superscript in parentheses $\left(m\right)$ indicates the $m^{th}$ realisation of the system, $u_{i}$ is the velocity, with subscripts representing the Einstein index convention, $p$ is the pressure, and $f_{i}$ a body force. The conformation tensor $c_{ij}$ is a macroscopic measure of the polymeric deformation. The dimensionless parameters controlling the flow are the Reynolds number $Re$, the Weissenberg number $Wi$, and the solvent to total viscosity ratio $\beta$. The polymers are assumed to obey a simplified linear PTT constitutive law~\citep{ptt_1977}, with the conformation tensor following an evolution equation given
\begin{equation}\frac{\partial{c}_{ij}^{\left(m\right)}}{\partial{t}}+u_{k}^{\left(m\right)}\frac{\partial{c}_{ij}^{\left(m\right)}}{\partial{x}_{k}}-\frac{\partial{u}_{i}^{\left(m\right)}}{\partial{x}_{k}}c_{kj}^{\left(m\right)}-\frac{\partial{u}_{j}^{\left(m\right)}}{\partial{x}_{k}}c_{ik}^{\left(m\right)}=-\frac{c_{ij}^{\left(m\right)}-\delta_{ij}}{Wi}\left(1-d\varepsilon+\varepsilon{c}_{kk}^{\left(m\right)}\right)+\kappa\frac{\partial^{2}c_{ij}^{\left(m\right)}}{\partial{x}_{k}\partial{x}_{k}},\label{eq:cte}\end{equation}
in which $\kappa$ is a dimensionless polymeric diffusivity, and $\varepsilon\ge0$ is the sPTT nonlinearity parameter. $d$ is the number of spatial dimensions. In the limit $\varepsilon=0$, the Oldroyd B constitutive equation is recovered. Note in steady homogeneous flows \hlb{the velocity and stress fields obtained with} the PTT model used here \hlb{are} identical to \hlb{those obtained with }the FENE-P model~\hlb{\mbox{\cite{davoodi_2022}}}.

Following~\cite{ge_2023} we consider two realisations of~\eqref{eq:mass} to~\eqref{eq:cte} ($m=1$ and $m=2$). For time $t<t_{0}$, these realisations are identical. At $t=t_{0}$ a small perturbation is imposed on realisation $\left(2\right)$, after which time the realisations may diverge. We define the difference between the realisations as $\Delta{u}_{i}=u_{i}^{\left(2\right)}-u_{i}^{\left(1\right)}$, $\Delta{p}=p^{\left(2\right)}-p^{\left(1\right)}$, and $\Delta{c}_{ij}=c_{ij}^{\left(2\right)}-c_{ij}^{\left(1\right)}$. We are interested in the evolution of these difference fields.

We can obtain evolution equations for $\Delta{u}_{i}$ and $\Delta{c}_{ij}$ by subtracting realisation $1$ from $2$ of~\eqref{eq:mass} to \eqref{eq:cte}. Doing so, expressing only in terms of realisation $1$ and the difference, and exploiting the symmetry of $c_{ij}^{\left(m\right)}$, we obtain
\begin{equation}\frac{\partial\Delta{u}_{i}}{\partial{x}_{i}}=0\label{eq:diffmass}\end{equation}
\begin{multline}\frac{\partial\Delta{u}_{i}}{\partial{t}}+u^{\left(1\right)}_{j}\frac{\partial\Delta{u}_{i}}{\partial{x}_{j}}+\Delta{u}_{j}\frac{\partial\Delta{u}_{i}}{\partial{x}_{j}}+\Delta{u}_{j}\frac{\partial{u}^{\left(1\right)}_{i}}{\partial{x}_{j}}=\\-\frac{\partial\Delta{p}}{\partial{x_{i}}}+\frac{\beta}{Re}\frac{\partial^{2}\Delta{u}_{i}}{\partial{x}_{j}\partial{x}_{j}}+\frac{1-\beta}{ReWi}\frac{\partial\Delta{c}_{ji}}{\partial{x}_{j}}+\Delta{f}_{i}\label{eq:diffmom}\end{multline}
\begin{multline}\frac{\partial\Delta{c}_{ij}}{\partial{t}}+u^{\left(1\right)}_{k}\frac{\partial\Delta{c}_{ij}}{\partial{x}_{k}}+\Delta{u}_{k}\frac{\partial\Delta{c}_{ij}}{\partial{x}_{k}}+\Delta{u}_{k}\frac{\partial{c}^{\left(1\right)}_{ij}}{\partial{x}_{k}}\\-\frac{\partial{u}^{\left(1\right)}_{i}}{\partial{x}_{k}}\Delta{c}_{kj}-\frac{\partial\Delta{u}_{i}}{\partial{x}_{k}}{c}^{\left(1\right)}_{kj}-\frac{\partial\Delta{u}_{i}}{\partial{x}_{k}}\Delta{c}_{kj}-\frac{\partial{u}^{\left(1\right)}_{j}}{\partial{x}_{k}}\Delta{c}_{ki}-\frac{\partial\Delta{u}_{j}}{\partial{x}_{k}}{c}^{\left(1\right)}_{ki}-\frac{\partial\Delta{u}_{j}}{\partial{x}_{k}}\Delta{c}_{ki}\\=-\frac{\Delta{c}_{ij}\left(1-d\varepsilon\right)}{Wi}+\frac{\delta_{ij}\varepsilon\Delta{c}_{kk}}{Wi}-\frac{\varepsilon}{Wi}\left(c_{ij}^{\left(1\right)}\Delta{c}_{kk}+\Delta{c}_{ij}c_{kk}^{\left(1\right)}+\Delta{c}_{ij}\Delta{c}_{kk}\right)\\+\kappa\frac{\partial^{2}\Delta{c}_{ij}}{\partial{x}_{k}\partial{x}_{k}}\label{eq:diffcte}.\end{multline}
In the present work we restrict consideration to flows with constant, flow-independent forcing, such that $\Delta{f}_{i}=0$, and for brevity of exposition, all terms containing $\Delta{f}_{i}$ are omitted hereafter. We note here that~\eqref{eq:diffmom} and~\eqref{eq:diffcte} show that a non-zero difference field in \emph{either} velocity \emph{or} conformation tensor may evolve into a non-zero difference in both fields. There is inherent uncertainty in the conformation tensor, being a statistical average of molecular deformations, and subject to (e.g.) thermal and concentration fluctuations.

\subsection{Uncertainty in the flow}
We define a positive scalar metric to quantify the \emph{uncertainty in the flow} as the kinetic energy of the velocity difference: $E_{\Delta}=\frac{1}{2}\Delta{u}_{i}\Delta{u}_{i}$. An evolution equation for $E_{\Delta}$ can be obtained by multiplying~\eqref{eq:diffmom} by $\Delta{u}_{i}$, resulting in
\begin{multline}\frac{\partial{E}_\Delta}{\partial{t}}+\frac{\partial\left(u_{j}^{\left(1\right)}+\Delta{u}_{j}\right)E_{\Delta}}{\partial{x}_{j}}+\Delta{u}_{i}\Delta{u}_{j}\frac{\partial{u}^{\left(1\right)}_{i}}{\partial{x}_{j}}=-\frac{\partial\Delta{u}_{i}\Delta{p}}{\partial{x_{i}}}+\\\frac{\beta}{Re}\frac{\partial}{\partial{x}_{j}}\left[\frac{\partial{E}_{\Delta}}{\partial{x}_{j}}\right]-\frac{\beta}{Re}\frac{\partial\Delta{u}_{i}}{\partial{x}_{j}}\frac{\partial\Delta{u}_{i}}{\partial{x}_{j}}+\frac{1-\beta}{ReWi}\Delta{u}_{i}\frac{\partial\Delta{c}_{ji}}{\partial{x}_{j}}.\label{eq:keu}\end{multline}
We take the average of~\eqref{eq:keu} over the periodic domain, and denoting the spatial average $\left\langle\cdot\right\rangle$ we obtain
\begin{equation}\frac{d\left\langle{E}_\Delta\right\rangle}{d{t}}=\left\langle-\Delta{u}_{i}S^{\left(1\right)}_{ij}\Delta{u}_{j}\right\rangle-\left\langle\frac{\beta}{Re}\frac{\partial\Delta{u}_{i}}{\partial{x}_{j}}\frac{\partial\Delta{u}_{i}}{\partial{x}_{j}}\right\rangle+\left\langle-\frac{1-\beta}{ReWi}\Delta{c}_{ij}\Delta{S}_{ij}\right\rangle\label{eq:akeu},\end{equation}
where $S^{\left(1\right)}_{ij}=\frac{1}{2}\left(\frac{\partial{u}_{i}^{\left(1\right)}}{\partial{x}_{j}}+\frac{\partial{u}_{j}^{\left(1\right)}}{\partial{x}_{i}}\right)$, and we have exploited the fact that $\left\langle\partial\left(\cdot\right)_{k}/\partial{x}_{k}\right\rangle=0$ on the periodic domain. Note that~\eqref{eq:akeu} is only valid on periodic domains; in more general settings, the evolution of $E_{\Delta}$ is given by~\eqref{eq:keu}. The first two terms on the right hand side of~\eqref{eq:akeu}, which (respectively) represent inertial production of uncertainty (denoted $I_{\Delta}$) and viscous dissipation of uncertainty (denoted $D_{\Delta}$) match those derived by~\cite{ge_2023} for Newtonian flows. The final term represents the polymeric \hla{propagation} of uncertainty (denoted $P_{\Delta}$)\hla{, and governs the propagation of uncertainty from the polymer deformation to the flow}. We express~\eqref{eq:akeu} in these terms as
\begin{equation}\frac{d\left\langle{E}_\Delta\right\rangle}{d{t}}=\left\langle{I}_{\Delta}\right\rangle-\left\langle{D}_{\Delta}\right\rangle+\left\langle{P}_{\Delta}\right\rangle.\end{equation}
Viscous dissipation $D_{\Delta}$ is positive by construction; Newtonian viscosity always reduces uncertainty. Inertial production may be positive or negative, and its sign is dependent on the alignment of the uncertainty velocity field $\Delta{u}_{i}$ with the reference strain rate, as discussed in detail in~\cite{ge_2023} in the context of three-dimensional inertial turbulence. Considering for now the two-dimensional case, we denote the eigenvalues of ${S}_{ij}^{\left(1\right)}$ as $\Lambda_{i}^{\left(1\right)}$ (following the convention that $\Lambda_{1}\le\Lambda_{2}$), and note that due to incompressibility the two eigenvalues are real, equal and opposite ($\Lambda_{2}^{\left(1\right)}=-\Lambda_{1}^{\left(1\right)}=\Lambda^{\left(1\right)}>0$). We define the components of $\Delta{u}_{i}$ projected onto the corresponding principal axes as $\Delta{u}_{i}^{\prime}$, and the inertial production term can be expressed as
\begin{equation}\left\langle{I}_{\Delta}\right\rangle=\left\langle\Lambda^{\left(1\right)}\left(\Delta{u}_{1}^{\prime{2}}-\Delta{u}_{2}^{\prime{2}}\right)\right\rangle,\end{equation}
in which $\Delta{u}_{1}^{\prime}$ corresponds to a velocity difference aligned with the compressive flow direction in the reference field, and $\Delta{u}_{2}^{\prime}$ corresponds to a velocity difference aligned with the stretching direction. This is the two-dimensional equivalent of the expression given by~\cite{ge_2023} for three-dimensional flows: the inertial production of uncertainty is increased by uncertainty aligned with compressive flows, and decreased by uncertainty aligned with flow stretching. A similar approach may be taken to provide insight into the polymeric \hla{propagation} of uncertainty $\left\langle{P}_{\Delta}\right\rangle$, which depends on the alignment of the uncertainty in the conformation tensor with the uncertainty in the strain rate. We again consider the two-dimensional case, and now take the principal axes of $\Delta{S}_{ij}$ as a local orthonormal reference frame. Denoting the eigenvalues of $\Delta{S}_{ij}$ as $\pm\Lambda_{\Delta}$ ($\Lambda_{\Delta}>0$), and the components of $\Delta{c}$ in this basis as $\Delta{c}_{ij}^{\star}$, we obtain
\begin{equation}\left\langle{P}^{2D}_{\Delta}\right\rangle=\left\langle\frac{1-\beta}{ReWi}\left[\left(\Delta{c}^{\star}_{11}-\Delta{c}^{\star}_{22}\right)\Lambda_{\Delta}\right]\right\rangle.\label{eq:pp2dp}\end{equation}
$\Delta{c}_{11}^{\star}$ corresponds to a difference in the polymer deformation aligned with compression in the velocity difference field, whilst $\Delta{c}_{22}^{\star}$ corresponds to a difference in the polymer deformation aligned with extension in the velocity difference field. We denote by $\Delta{N}_{1}^{\star}$ the term in parentheses, which is proportional to the difference in the first normal stress difference in a basis aligned with a compressive flow difference. The polymeric \hla{propagation} of uncertainty depends on the difference in the first normal stress difference in a basis aligned with the principal axes of the velocity difference. Little can be said a-priori of the likely sign or magnitude of $\Delta{N}_{1}^{\star}$, as $\Delta{c}_{ij}$ is not constrained to remain positive definite, and even if both realisations exhibit a positive first normal stress the difference, $\Delta{N}_{1}^{\star}$ may be negative. 

Considering the viscous dissipation of uncertainty $\left\langle{D}_{\Delta}\right\rangle$, in two dimensions we may write
\begin{equation}D_{\Delta}=\frac{\beta}{Re}\frac{\partial\Delta{u}_{i}}{\partial{x}_{j}}\frac{\partial\Delta{u}_{i}}{\partial{x}_{j}}=\frac{\beta}{Re}\left(\Delta{S}_{ij}\Delta{S}_{ij}+\frac{1}{2}\Delta\omega^{2}\right),\end{equation}
where $\Delta\omega$ is the vorticity difference. As before, we can take the principal axes of $\Delta{S}_{ij}$ as a local orthonormal basis, and we obtain
\begin{equation}D_{\Delta}=\frac{\beta}{Re}\left(2\Lambda_{\Delta}^{2}+\frac{1}{2}\Delta\omega^{2}\right),\end{equation}
highlighting that viscous dissipation occurs due to both differences in extension ($\Lambda_{\Delta}$) and vorticity ($\Delta\omega$).

%\subsubsection{Symmetry}
%If write evolution equations for $\Delta{u}_{i}$, $E_{\Delta}$ and $\left\langle{E}_{\Delta}\right\rangle$ in terms of the perturbed field rather than the reference field, we obtain for $\left\langle{E}_{\Delta}\right\rangle$
%\begin{equation}\frac{d\left\langle{E}_\Delta\right\rangle}{d{t}}=\left\langle-\Delta{u}_{i}S^{\left(2\right)}_{ij}\Delta{u}_{j}\right\rangle-\left\langle\frac{\beta}{Re}\frac{\partial\Delta{u}_{i}}{\partial{x}_{j}}\frac{\partial\Delta{u}_{i}}{\partial{x}_{j}}\right\rangle+\left\langle-\frac{1-\beta}{ReWi}\Delta{c}_{ij}\Delta{S}_{ij}\right\rangle\label{eq:akeu2}.\end{equation}
%Note that the strain rate tensor appearing in the advective term is now that of the perturbed field. Hence
%\begin{equation}\left\langle-\Delta{u}_{i}S^{\left(1\right)}_{ij}\Delta{u}_{j}\right\rangle=\left\langle-\Delta{u}_{i}S^{\left(2\right)}_{ij}\Delta{u}_{j}\right\rangle,\end{equation}
%which may be useful. Note the equilance is not true locally (i.e. without the spatial averaging).

As a consequence of the elliptic nature of the incompressibility constraint in~\eqref{eq:mass} and~\eqref{eq:mom}, a perturbation to the flow which is local in space can generate a non-zero uncertainty over the entire flow field instantly. For uncertainty in the flow this non-local behaviour is via the pressure gradient term in~\eqref{eq:diffmom}. The pressure-gradient term in~\eqref{eq:keu} disappears on spatial averaging (i.e. it does not appear in~\eqref{eq:akeu}). Whilst it therefore does not explicitly contribute to the evolution of $\left\langle{E}_{\Delta}\right\rangle$ as a global measure of uncertainty, it does still influence the evolution of $\left\langle{E}_{\Delta}\right\rangle$, through its influence on both $\Delta{u}_{i}$ and $\Delta{c}_{ii}$. This is verified by numerical experiments, and discussed further in~\S~\ref{ne}. 

\subsection{Uncertainty in polymer deformation}

Identifying a suitable scalar measure of the uncertainty in the conformation tensor is non-trivial. Two requirements for such a measure are i) that it is strictly positive, and ii) that it is independent of the conformation tensor basis. Although $c_{ij}^{\left(1\right)}$ and $c_{ij}^{\left(2\right)}$ are (symmetric) positive definite, the invariants of $\Delta{c}_{ij}$ are not necessarily positive. Several approaches are offered by~\cite{hameduddin_2018}, including the logarithmic volume ratio, and the squared distance between $c_{ij}^{\left(1\right)}$ and $c_{ij}^{\left(2\right)}$ along a geodesic curve on a Riemannian manifold. Whilst both approaches provide a strictly positive scalar measure, the derivation of evolution equations for these measures becomes prohibitively complex. For the Oldroyd B model, the elastic energy stored in the polymers is proportional to the conformation tensor trace, and a measure which can be related to the uncertainty in the stored elastic energy is desirable. We opt to take the square of the conformation tensor trace difference as a scalar measure of uncertainty on the polymeric deformation, which satisfies both the above requirements. We define $\Gamma_{\Delta}=\left(\Delta{c}_{ii}\right)^{2}$, multiply the trace of~\eqref{eq:diffcte} by $2\Delta{c}_{jj}$, to obtain an evolution equation for $\Gamma_{\Delta}$:
\begin{multline}\frac{\partial\Gamma_{\Delta}}{\partial{t}}+u^{\left(1\right)}_{k}\frac{\partial\Gamma_{\Delta}}{\partial{x}_{k}}+\Delta{u}_{k}\frac{\partial\Gamma_{\Delta}}{\partial{x}_{k}}+2\Delta{c}_{jj}\Delta{u}_{k}\frac{\partial{c}^{\left(1\right)}_{ii}}{\partial{x}_{k}}\\-4\Delta{c}_{jj}\Delta{c}_{ik}S_{ik}^{\left(1\right)}-4\Delta{c}_{jj}{c}^{\left(1\right)}_{ik}\Delta{S}_{ik}-4\Delta{c}_{jj}\Delta{c}_{ik}\Delta{S}_{ik}\\=-\frac{2\Gamma_{\Delta}}{Wi}\left[1+2\varepsilon{c}_{ii}^{\left(1\right)}+\varepsilon\Delta{c}_{ii}\right]+\kappa\frac{\partial^{2}\Gamma_{\Delta}}{\partial{x}_{k}\partial{x}_{k}}-{2}\kappa\frac{\partial\Delta{c}_{ii}}{\partial{x}_{k}}\frac{\partial\Delta{c}_{jj}}{\partial{x}_{k}}.\end{multline}
We take the spatial average, and again exploiting the fact that any terms expressed in flux form average to zero over a periodic domain, we obtain
\begin{multline}\frac{d\left\langle\Gamma_{\Delta}\right\rangle}{dt}=\left\langle-2\Delta{c}_{ii}\Delta{u}_{k}\frac{\partial{c}^{\left(1\right)}_{jj}}{\partial{x}_{k}}\right\rangle+\left\langle4\Delta{c}_{jj}\Delta{c}_{ik}S_{ik}^{\left(1\right)}\right\rangle+\left\langle4\Delta{c}_{jj}{c}^{\left(1\right)}_{ik}\Delta{S}_{ik}\right\rangle+\left\langle4\Delta{c}_{jj}\Delta{c}_{ik}\Delta{S}_{ik}\right\rangle\\-\left\langle\frac{2\Gamma_{\Delta}}{Wi}\left[1+2\varepsilon{c}_{ii}^{\left(1\right)}+\varepsilon\Delta{c}_{ii}\right]\right\rangle-\left\langle{2}\kappa\frac{\partial\Delta{c}_{ii}}{\partial{x}_{k}}\frac{\partial\Delta{c}_{jj}}{\partial{x}_{k}}\right\rangle.\label{eq:dG}\end{multline}
The first term on the right hand side relates to the production of uncertainty due to advection. The next three terms originate from the upper convected derivative, and relate to \hla{the} production \hla{and propagation} of uncertainty due to stretching and rotation of the polymers. The penultimate term corresponds to the \hla{destruction} of uncertainty due to polymer relaxation. The final term is the dissipation of uncertainty due to polymeric diffusion. For convenience, we will re-write~\eqref{eq:dG} with the terms denoted
\begin{equation}\frac{d\left\langle\Gamma_{\Delta}\right\rangle}{dt}=\left\langle{A}_{\Delta}\right\rangle+\left\langle{UC1}_{\Delta}\right\rangle+\left\langle{UC2}_{\Delta}\right\rangle+\left\langle{UC3}_{\Delta}\right\rangle-\left\langle{R}_{\Delta}\right\rangle-\left\langle{PD}_{\Delta}\right\rangle.\end{equation}
We first comment on the effect of polymer relaxation on uncertainty. In the limit of $\varepsilon\to0$ (the Oldroyd B limit) $R_{\Delta}$ is positive by construction; in the Oldroyd B limit, polymer relaxation always reduces uncertainty in the polymer deformation. For finite $\varepsilon$, the non-linearity in the polymer relaxation may result in (locally) negative $R_{\Delta}$, when and where the inequality $c_{ii}^{\left(2\right)}>3c_{ii}^{\left(1\right)}+1/\varepsilon$ is satisfied. Certainly in the early stages of uncertainty evolution, when the two fields are closely correlated, we would not expect this to occur. With the two realisations following the same evolution equations, and expected to have the same statistics, we would not expect $\left\langle{R}_{\Delta}\right\rangle<0$ either. Indeed, $\left\langle{R}_{\Delta}\right\rangle$ remains positive throughout all our numerical simulations, and in fact we see a trend that $\left\langle{R}_{\Delta}\right\rangle/\left\langle\Gamma_{\Delta}\right\rangle$ increases with increasing $\varepsilon$. The term $\left\langle{PD}_{\Delta}\right\rangle$ is positive by construction; polymeric diffusion always reduces uncertainty in the polymer deformation. The \hla{remaining terms} $\left\langle{A}_{\Delta}\right\rangle$, $\left\langle{UC1}_{\Delta}\right\rangle$, $\left\langle{UC2}_{\Delta}\right\rangle$, and $\left\langle{UC3}_{\Delta}\right\rangle$, may be either positive or negative. \hlb{We note that an alternative positive scalar measure of the uncertainty in the polymer deformation could be defined as $\Pi_{\Delta}=\Delta{c}_{ij}\Delta{c}_{ij}$. This measure is frame invariant, and includes information on the off-diagonal elements of $\Delta{c}_{ij}$; note, in two dimensions, we can write $\Pi_{\Delta}=\Gamma_{\Delta}-2\det\left(\Delta{c}_{ij}\right)$. Although we do not explore this measure in detail, we include it in our investigation of the orientation of uncertainty in\S~\mbox{\ref{sec:orient}}, and an evolution equation for $\Pi_{\Delta}$ is provided in Appendix~\mbox{\ref{sec:pi}}.}

\hlb{Returning to the evolution of $\left\langle\Gamma_{\Delta}\right\rangle$ and c}onsidering the \hla{production} term $\left\langle{UC1}_{\Delta}\right\rangle$, we take the principal axes of ${S}_{ij}^{\left(1\right)}$ as a local orthonormal basis, and we denote the components of $\Delta{c}$ in this basis as $\Delta{c}_{ij}^{\prime}$. Expressed in this basis, we can write
\begin{equation}UC1_{\Delta}=4\Lambda^{\left(1\right)}\left(\Delta{c}_{22}^{\prime{2}}-\Delta{c}_{11}^{\prime{2}}\right)\label{eq:uc1},\end{equation}
where $\Delta{c}_{11}^{\prime}$ represents deformation in the conformation tensor difference aligned with compressive flow in the reference field, and $\Delta{c}_{22}^{\prime}$ represents that aligned with stretching flow. From~\eqref{eq:uc1}, we see that regions of extensional deformation in the conformation tensor difference aligned with extensional flow \hla{produce} uncertainty, whilst extensional deformation in the conformation tensor difference aligned with compressive flow reduce uncertainty. Next considering the term $\left\langle{UC2}_{\Delta}\right\rangle$, \hla{which is a propagation or transfer term, through which uncertainty in the flow (via $\Lambda_{\Delta}$) propagates into the conformation tensor uncertainty (but only where some uncertainty in the polymeric deformation already exists i.e. $\Delta{c}_{ii}\ne0$). We} tak\hla{e} the principal axes of $\Delta{S}_{ij}$ as a local orthornormal basis, \hla{and} write
\begin{equation}UC2_{\Delta}=4\Delta{c}_{ii}\Lambda_{\Delta}\left(c_{22}^{\left(1\right)\star}-c_{11}^{\left(1\right)\star}\right),\label{eq:uc2}\end{equation}
in which $c_{22}^{\left(1\right)\star}$ and $c_{11}^{\left(1\right)\star}$ correspond to polymeric deformation in the reference field aligned with stretching and compression (respectively) in the velocity difference field. Whether or not this term is positive or negative depends on the sign of the conformation tensor trace difference, and the normal stress difference in the reference field aligned with the principal stretching direction of the velocity difference. The term $\left\langle{UC3}_{\Delta}\right\rangle$\hla{, also describing uncertainty propagation,} is constructed from the product of three differences, whilst the terms $\left\langle{UC1}_{\Delta}\right\rangle$ and $\left\langle{UC2}_{\Delta}\right\rangle$ are each formed of the product of two differences and a reference field. Hence, whilst the two realisations are largely correlated and uncertainty is small, we expect the term $\left\langle{UC3}_{\Delta}\right\rangle$ to be small relative to the other two upper convected terms. The term in~\eqref{eq:dG} relating to the advection of polymers contains only invariants of the conformation tensor and its difference. If we take the principal axes of $c_{ij}^{\left(1\right)}$ as a local orthonormal basis, and denote tensors expressed in this basis, and vectors projected onto this basis, with a double prime, we can express the advection term in~\eqref{eq:dG} as
\begin{equation}{A}_{\Delta}=-2\Delta{c}_{ii}\Delta{u}^{\prime\prime}_{k}\frac{\partial{c}_{jj}^{\left(1\right)}}{\partial{x}^{\prime\prime}_{k}}.\end{equation}
A characteristic of elastic turbulence is thin regions of highly deformed polymers, and here the gradient of the conformation tensor trace $\partial{c}^{\left(1\right)}_{ii}/\partial{x}_{k}$ will be predominantly aligned normal to the principal stretching direction of the polymers. Hence, we expect $\partial{c}_{jj}^{\left(1\right)}/\partial{x}^{\prime\prime}_{1}\gg\partial{c}_{jj}^{\left(1\right)}/\partial{x}^{\prime\prime}_{2}$ in regions of large polymer deformation. Noting that $\Delta{c}_{ii}$ may be negative, but whilst the two realisations are still predominantly correlated, $\left\lvert\Delta{c}_{ii}\right\rvert$ is likely to be larger in regions where $c_{ii}^{\left(1\right)}$ is large, we may expect $\left\langle{A}_{\Delta}\right\rangle$ to depend on the relative orientation of the uncertainty in the velocity, and the reference polymer deformation. Such analysis can only provide a certain amount of insight, and to investigate the dynamics of uncertainty further, we turn to numerical simulations. 

\section{Numerical Experiments}\label{ne}

We conduct numerical experiments on the system described in~\S\ref{ta}. We consider two-dimensional realisations of~\eqref{eq:mass} to~\eqref{eq:cte} on a doubly-periodic domain with (integer) side length $n$, subject to a constant cellular forcing. The system is non-dimensionalised by the forcing wavelength and the velocity magnitude of the Newtonian laminar fixed point, and setting the forcing $f_{1}=f_{0}\sin\left(2\pi{x}_{2}\right)$, $f_{2}=f_{0}\sin\left(2\pi{x}_{1}\right)$, with $f_{0}=4\pi^{2}/Re$ gives, for the Newtonian case ($\beta\to{1}$ or $Wi\to{0}$) at small $Re$, a stable laminar fixed point $u_{1}=\sin\left(2\pi{x}_{2}\right)$, $u_{2}=\sin\left(2\pi{x}_{2}\right)$. We note that this configuration is closely related to those studied in~\cite{plan_2017}, but with a different non-dimensionalisation: we take the forcing scale as the characteristic lengthscale, and impose a forcing magnitude inversely proportional to $Re$, whilst~\cite{plan_2017} take the domain size as the characteristic length scale, and do not provide information on the relationship between $f_{0}$ and $Re$.

We numerically solve~\eqref{eq:mass} to~\eqref{eq:cte} using a pseudo-spectral code built within the Dedalus framework~\citep{burns_2020}. The domain is discretised with $128$ Fourier modes per forcing wavelength (i.e. the entire domain is discretised with $\left(128n\right)^{2}$) unless specified otherwise. The system is integrated in time with a $2^{nd}$ order Runge-Kutta scheme, and except where explicitly stated, we use a fixed time step of $\delta{t}=4\times{10}^{-4}$. We base our experiments around a reference configuration with $Re=10^{-2}$, $\beta=1/2$, $\varepsilon=0$, $\kappa=2.5\times{10}^{-5}$, $n=4$, $Wi=2$. \hlc{Whilst for our primary investigation we set $\varepsilon=0$ to obtain the Oldroyd B model, we include a study of the effects of the sPTT nonlinearity parameter in Appendix~\mbox{\ref{sec:ptt}}.} We include a polymeric diffusivity term in our simulations to provide numerical stability. \hla{The topic of polymeric diffusivity is one of debate within the community, with many early simulations of chaotic polymer flows using values of diffusivity much larger than the physical values for real polymer solutions (see~\mbox{\cite{dubief_2023}} for a discussion on this topic).} As explored in detail in~\cite{gupta_2019}, care must be taken when including polymeric diffusivity in a numerical simulation. \hla{An argument can be made that there is a physical basis for the inclusion of polymeric diffusivity, and several researchers (e.g.~\mbox{\cite{morozov_2022,nichols_2025}}) justify the use of} polymeric diffusivity by kinetic theory\hla{. A rigorous justification remains elusive, and there are open questions about whether the polymeric diffusivity used in numerical simulations is modelled appropriately form (e.g. is real polymer diffusivity really spatially uniform and flow/polymer deformation independent?).} The value \hla{of $\kappa$} we use is slightly smaller than th\hla{at used in~\mbox{\cite{morozov_2022}}}. \hla{The values of $\kappa$ and $Re$ in our reference state above correspond to a Schmidt number of $Sc=4\times{10}^{6}$, and this value is in agreement with the physical value of $Sc$ for polymers in water suggested by~\mbox{\cite{dubief_2023}}. However, we again highlight that our decision to include polymeric diffusivity is based on numerical necessity.} We note that whilst~\cite{morozov_2022} was exploring the onset of instabilities, we are investigating the dynamics of an established chaotic state. Consequently, \hla{in contrast to the simulations presented by~\mbox{\cite{morozov_2022}},} our results are not independent of $\kappa$. Due to the chaotic nature of the flow, and the sensitivity to perturbations which we will explore in the following sections, this dependence on polymeric diffusivity is expected: for a chaotic flow, we expect the evolution of two realisations with the same initial conditions and differing only by a small change in the polymeric diffusivity $\Delta\kappa$ to diverge over time. \hlb{We discuss this dependence in more detail in~\S\mbox{\ref{sec:sc_nl}}.} 

\begin{figure}
\includegraphics[width=0.49\textwidth]{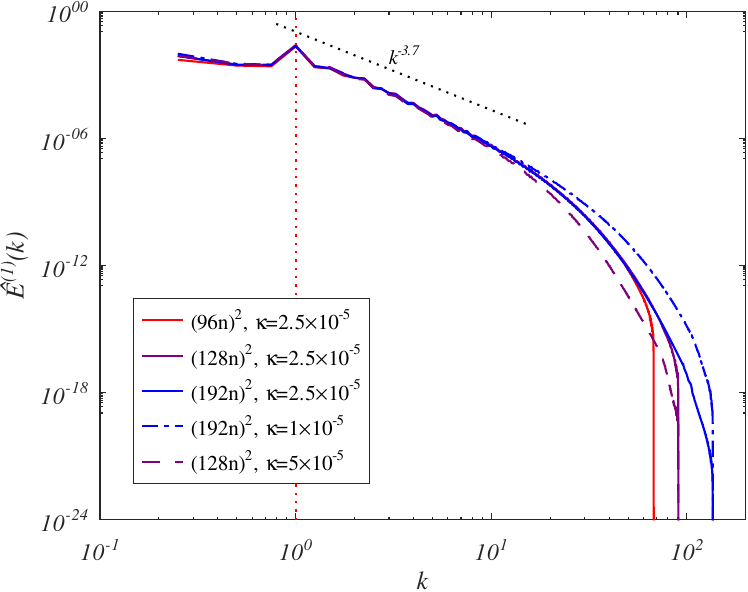}
\includegraphics[width=0.45\textwidth]{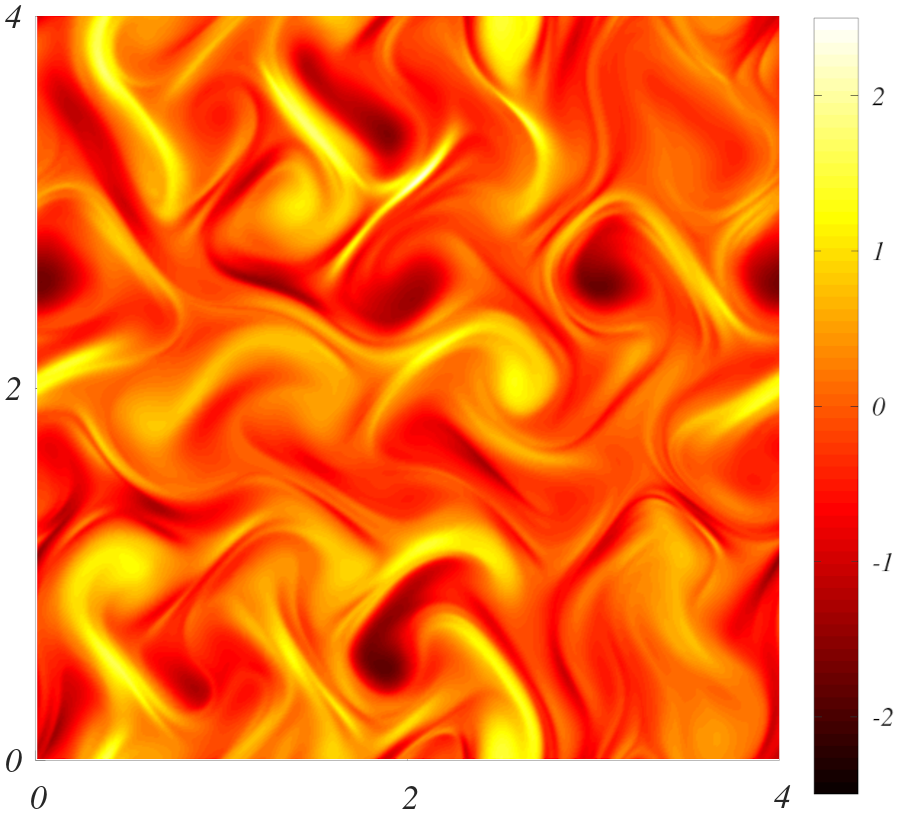}
\caption{Left panel: The kinetic energy spectra of the flow in the reference configuration, for \hla{three} resolutions \hla{$\left(96n\right)^{2}$,} $\left(128n\right)^{2}$ and $\left(192n\right)^{2}$\hla{ (solid lines), and for lower ($\kappa=5\times10^{-5}$, dash-dot line) and higher ($\kappa=10^{-5}$, dashed line) polymeric diffusivity}. Right panel: a snapshot of the vorticity field for the reference configuration. \label{fig:ref_conv}}
\end{figure}

Additional simulations of the reference configuration at \hla{lower ($\left(96n\right)^{2}$ modes) and }higher resolution\hla{s} ($\left(192n\right)^{2}$ modes) confirm resolution independence. The simulations are run for $200$ time units to allow initial transients to decay. For the largest domain sizes and Weissenberg numbers considered, this is more than $10$ times the maximum timescale of the flow. Statistics are then gathered for a further $200$ time units. Between $\left(128n\right)^{2}$ and $\left(192n\right)^{2}$ modes the  discrepancy in the mean kinetic energy is $0.35\%$, and the standard deviation of the kinetic energy differs by $1.6\%$. In the Oldroyd B limit ($\varepsilon=0$), there are lower bounds on the conformation tensor trace ($c_{ii}\ge{2}$ in two-dimensions) and its determinant ($\det{c}\ge{1}$)~\citep{hu_2007}: values below these bounds represent non-physical configurations of polymer deformation, and it has recently been shown by~\cite{yerasi_2024} that failing to satisfy these constraints can result in a numerical simulation exhibiting significantly modified large scale flow dynamics. Although these criteria are not enforced by construction in our numerical framework, we confirm that the resolution we use \hla{($\left(128n\right)^{2}$ modes)} is sufficient that both are met, \hla{whilst noting that at the lower resolution of $\left(96n\right)^{2}$ modes, these criteria are not always satisfied}. The left panel of figure~\ref{fig:ref_conv} shows the energy spectra of the flow for \hla{three} resolutions. Note that here, and in all other spectra calculated in this work, the wavenumbers have been normalised by the forcing wavenumber (indicated by a vertical dotted red line). There is a close agreement in the energy spectra up to the Nyquist wavenumber of the most coarse resolution. The energy spectra shows a power law decay, with a slope of approximately $3.7$ (indicated in figure~\ref{fig:ref_conv} by a dashed black line), characteristic of the elastic turbulent regime. \hlb{We also show in figure~\mbox{\ref{fig:ref_conv}} the spectra at lower (dashed line) and higher (dash-dot line) diffusivities, and reiterate here that our simulations are not independent of $\kappa$.} \hla{Further discussion and results demonstrating temporal- and spatial-resolution independence are provided in Appendix~\mbox{\ref{sec:conv}}.} The right panel of figure~\ref{fig:ref_conv} shows a snapshot of the vorticity field for the reference configuration.

For each configuration we run a precursor simulation for \hlc{$400$} time units (realisation $1$). Realisation $2$ is initialised by restarting the simulation from realisation $1$ at a specified time $t_{0}>200$, and subject to a small perturbation. We then track the evolution of the difference between the two realisations. We denote the time after the perturbation is imposed as $\tau=t-t_{0}$. The perturbation is imposed by the addition of the term 
\begin{equation}-A_{0}\exp\left[-\left(\frac{\tau}{\delta_{t0}}\right)^{2}\right]\frac{\partial^{4}c_{ij}^{\left(2\right)}}{\partial{x}_{k}\partial{x}_{k}\partial{x}_{l}\partial{x}_{l}}\qquad\forall\tau\ge0\label{eq:ptbn}\end{equation}
to the right hand side of~\eqref{eq:cte}. The parameter $\delta_{t0}$ controls the temporal extent of the perturbation, and we use $\delta_{t0}=10^{-3}$ throughout. $A_{0}$ controls the magnitude of the perturbation, and except where explicitly stated, we set $A_{0}=2.5\times{10}^{-8}$. This perturbation acts directly on the conformation tensor, and suppresses high wavenumber modes. We have investigated reducing $\delta_{t0}$, such that $\delta_{t0}\ll\delta{t}$, where $\delta{t}$ is the time step used in the numerical framework, and find that although the magnitude of the uncertainty is reduced (the integral of the term in~\eqref{eq:ptbn} is smaller) this does not influence the evolution of uncertainty.

The temporal evolution of the spatially averaged quantities exhibits a dependence on the reference flow. Hence, for each configuration we conduct $10$ simulations for $t_{0}\in\left[200,210,220,230,240,250,260,270,280,290\right]$. Simulations are run until $\tau=50$, allowing the uncertainty to reach a statistically steady state for the reference configuration. Except where stated in the following, when plotting the time evolution of spatially averaged quantities, we show the ensemble average across these $10$ simulations. For strictly positive spatially averaged quantities which evolve over many orders of magnitude (e.g. $\left\langle{E}_{\Delta}\right\rangle$ and $\left\langle\Gamma_{\Delta}\right\rangle$), the ensemble average is taken as the geometric mean. For normalisation purposes, we define an average total energy as
\begin{equation}E_{avg}^{\left(tot\right)}=\frac{1}{\tau_{0}}\displaystyle\int_{0}^{\tau_{0}}\left\langle{E}^{\left(1\right)}\right\rangle+\left\langle{E}^{\left(2\right)}\right\rangle{d}\tau,\end{equation}
where $\tau_{0}=50$ time units, the duration of the simulation.

\subsection{The evolution of uncertainty}

\begin{figure}
\includegraphics[width=0.49\textwidth]{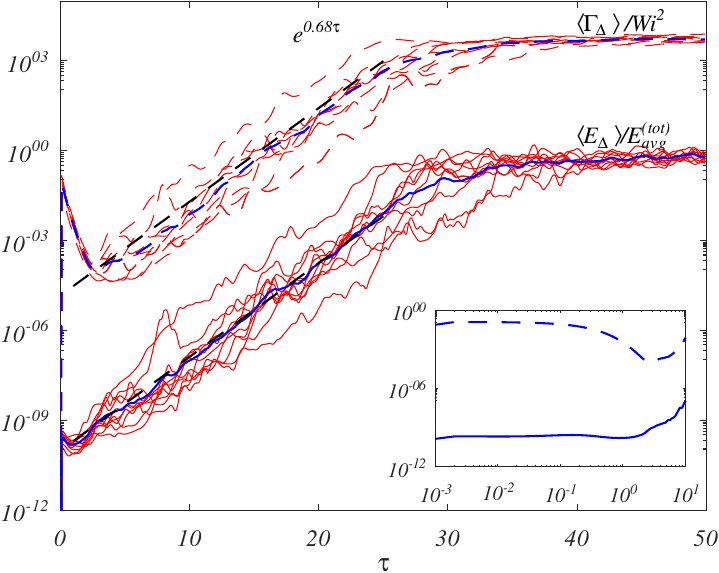}
\includegraphics[width=0.49\textwidth]{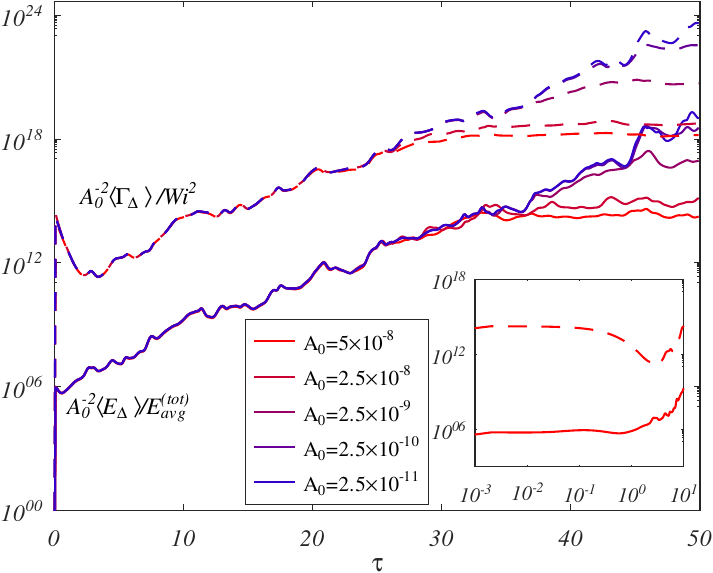}
\caption{Left panel: the evolution of $\left\langle{E}_{\Delta}\right\rangle/E_{avg}^{\left(tot\right)}$ and $\left\langle\Gamma_{\Delta}\right\rangle/Wi^{2}$ for the reference configuration ($Re=10^{-2}$, $\beta=1/2$, $\varepsilon=0$, $\kappa=2.5\times{10}^{-5}$, $n=4$, $Wi=2$). Red lines indicate the evolution for individual simulations, whilst blue lines indicate the average. Right panel: the evolution of $A_{0}^{-2}\left\langle{E}_{\Delta}\right\rangle/E_{avg}^{\left(tot\right)}$ and $A_{0}^{-2}\left\langle\Gamma_{\Delta}\right\rangle/Wi^{2}$ for a range of values of $A_{0}$. In both panels, the inset shows the evolution of uncertainty at small $\tau$.\label{fig:dEGref}}
\end{figure}

The left panel of figure~\ref{fig:dEGref} shows the evolution of $\left\langle{E}_{\Delta}\right\rangle/E_{avg}^{\left(tot\right)}$ and $\left\langle\Gamma_{\Delta}\right\rangle/Wi^{2}$ with $\tau$, for individual instances (red lines), and the ensemble average (blue lines). There is a clear period of exponential growth, over approximately $6$ orders of magnitude, with the same exponent in both $\left\langle{E}_{\Delta}\right\rangle/E_{avg}^{\left(tot\right)}$ and $\left\langle\Gamma_{\Delta}\right\rangle/Wi^{2}$. This exponential growth starts at $\tau\approx{1}$ for the flow uncertainty, and $\tau\approx3$ for $\left\langle{\Gamma}_{\Delta}\right\rangle$. As observed by~\cite{berti_2008,berti_2010} for elastic turbulence driven by a Kolmogorov forcing, there is a strong imprint of the laminar fixed point in the chaotic flow field, and as a consequence there is a limit to the uncertainty (the extent to which the two realisations can decorrelate). This is clear in the left panel of figure~\ref{fig:dEGref} as both $\left\langle{E}_{\Delta}\right\rangle/E_{avg}^{\left(tot\right)}$ and $\left\langle\Gamma_{\Delta}\right\rangle/Wi^{2}$ reach limiting values of approximately $0.3$ and $2000$ (respectively) after approximately $30$ time units. Note that $\left\langle\left({c}_{ii}^{\left(1\right)}\right)^{2}+\left({c}_{ii}^{\left(2\right)}\right)^{2}\right\rangle/Wi^{2}\approx6645$, and the saturation value of $\left\langle\Gamma_{\Delta}\right\rangle/Wi^{2}$ is about $0.3$ times this. From these results we can identify four regimes: (I) very short time growth $\left\langle{E}_{\Delta}\right\rangle$ and $\left\langle\Gamma_{\Delta}\right\rangle$ until $\tau\approx{0.1}$; (II) a reduction in uncertainty, initially in $\left\langle\Gamma_{\Delta}\right\rangle$, over timescales of the order of unity; (III) exponential growth of uncertainty, initially in $\left\langle{E}_{\Delta}\right\rangle$ from $\tau\approx{1}$, then also $\left\langle\Gamma_{\Delta}\right\rangle$ from $\tau\approx{4}$; and (IV) saturation of uncertainty from $\tau\approx{30}$. We discuss the mechanisms controlling these regimes further below. Note, we have run simulations for an additional $150$ time units, up to $\tau=200$, and do not observe any further growth in uncertainty.  During the exponential regime (III), the evolution of the quantities $\left\langle{E}_{\Delta}\right\rangle$ and $\left\langle\Gamma_{\Delta}\right\rangle$ can be modelled by
\begin{equation}\frac{d\left\langle{E}_{\Delta}\right\rangle}{dt}=\lambda\left\langle{E}_{\Delta}\right\rangle;\qquad\frac{d\left\langle{\Gamma}_{\Delta}\right\rangle}{dt}=\lambda\left\langle{\Gamma}_{\Delta}\right\rangle\end{equation}
in which $\lambda$ is a characteristic growth rate, equal to twice the maximal Lyapunov exponent. Letting $\left\{\cdot\right\}_{\left(III\right)}$ represent the temporal mean of a quantity over regime (III), we calculate $\left\{\lambda\right\}_{\left(III\right)}=0.68$. This growth rate is shown by the dashed black lines in the left panel of figure~\ref{fig:dEGref}. 

The right panel of figure~\ref{fig:dEGref} shows the evolution of $\left\langle{E}_{\Delta}\right\rangle/E_{avg}^{\left(tot\right)}$ with $\tau$ for different values of $A_{0}$, with results scaled by $A_{0}^{-2}$, for a single simulation. Prior to the saturation of uncertainty, the lines collapse, as the magnitude of the uncertainty scales with $A_{0}^{2}$, the square of the perturbation amplitude, by definition of $E_{\Delta}$ and $\Gamma_{\Delta}$. Just as the evolution of uncertainty is independent of $\delta_{t0}$ it is unchanged by the magnitude of $A_{0}$, despite the magnitude of the perturbation changing by over three orders of magnitude. The imprint of the reference flow on the uncertainty is visible \hlc{in} the right panel of figure~\ref{fig:dEGref}, where for all values of $A_{0}$ the same reference flow is used. 

\subsubsection{The orientation of uncertainty}\label{sec:orient}

As before, we take the principal axes of the reference strain rate tensor $S_{ij}^{\left(1\right)}$ to define a local orthonormal basis (denoted with a prime). We also use the principal axes of the reference conformation tensor $c_{ij}^{\left(1\right)}$ to define another basis (denoted with a double prime). We use each basis to obtain a projection of the flow uncertainty $\Delta{u}_{i}$, which we denote $\Delta{u}_{i}^{\prime}$ and $\Delta{u}_{i}^{\prime\prime}$. We also express the uncertainty in the conformation tensor in these bases (denoted $\Delta{c}_{ij}^{\prime}$ and $\Delta{c}_{ij}^{\prime\prime}$). We denote the proportion of the uncertainty energy aligned with each principal direction as
\begin{equation}\theta^{\prime}_{i}=\frac{\Delta{u}_{i}^{\prime{2}}}{2\left\langle{E}_{\Delta}\right\rangle},\end{equation}
where an equivalent definition follows for $\theta_{i}^{\prime\prime}$. Note that $\theta_{1}^{\prime}+\theta_{2}^{\prime}=\theta_{1}^{\prime\prime}+\theta_{2}^{\prime\prime}=1$. The quantity $\theta_{1}^{\prime}$ represents the proportion of uncertainty energy aligned with compressive flow and $\theta_{2}^{\prime}$ represents the proportion aligned with stretching flow. The quantity $\theta_{2}^{\prime\prime}$ represents the proportion of uncertainty energy aligned with the principal stretching direction of the conformation tensor, and $\theta_{1}^{\prime\prime}$ is the proportion of uncertainty energy orthogonal to this.

\begin{figure}
\includegraphics[width=0.49\textwidth]{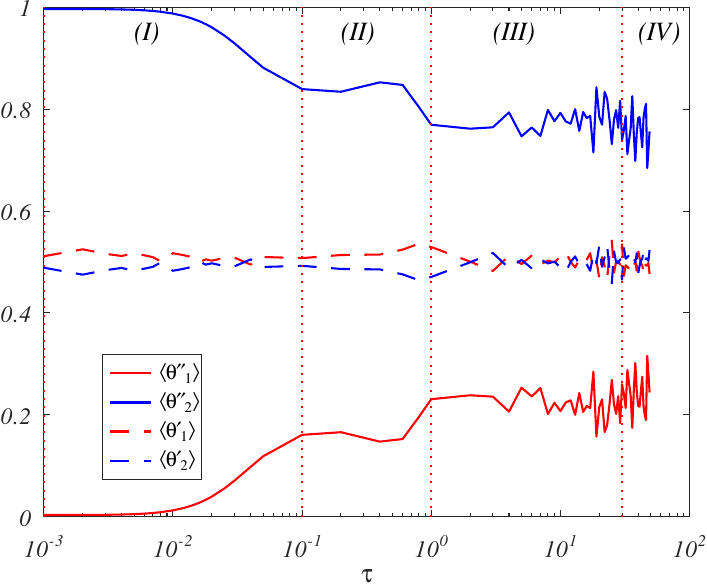}
\includegraphics[width=0.49\textwidth]{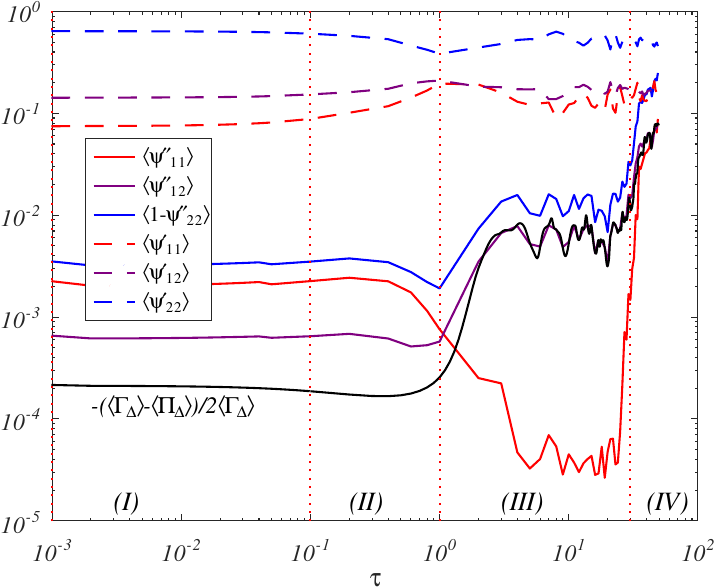}
\caption{The evolution of the orientation of uncertainty for the reference configuration ($Re=10^{-2}$, $\beta=1/2$, $\varepsilon=0$, $\kappa=2.5\times{10}^{-5}$, $n=4$, $Wi=2$). Left panel: the evolution of $\left\langle\theta^{\prime}\right\rangle$ and $\left\langle\theta^{\prime\prime}\right\rangle$. Right panel: the evolution of components of $\left\langle\psi^{\prime}\right\rangle$ and $\left\langle\psi^{\prime\prime}\right\rangle$. Note the components of $\left\langle\psi^{\prime\prime}\right\rangle$ may be greater than unity, and where we have plotted $\left\langle1-\psi_{22}^{\prime\prime}\right\rangle$ and this quantity is negative, it is plotted with a dotted line. \hlb{The black line shows the evolution of $-\left(\left\langle\Gamma_{\Delta}\right\rangle-\left\langle\Pi_{\Delta}\right\rangle\right)/2\left\langle\Gamma_{\Delta}\right\rangle$.} The different regimes of uncertainty evolution are indicated separated by dotted red vertical lines.\label{fig:pa}}
\end{figure}

In~\cite{ge_2023}, they observed an uneven distribution of the orientation of the uncertainty energy for three-dimensional turbulence in a Newtonian fluid: during the similarity regime characterised by exponential growth of uncertainty, the majority of the uncertainty energy was aligned with the compressive flow direction. The picture is different in the two-dimensional low $Re$ elastic turbulence case. The left panel of figure~\ref{fig:pa} shows the evolution of $\left\langle\theta^{\prime}\right\rangle$ and $\left\langle\theta^{\prime\prime}\right\rangle$. The variation of $\left\langle\theta_{1}^{\prime}\right\rangle$ is small: at all times $0.467\le\left\langle\theta_{1}^{\prime}\right\rangle\le0.542$; the uncertainty energy is approximately evenly split between regions of compression and extension. There does not appear to be any preferential orientation of uncertainty energy due to the reference flow. The relative orientation of the velocity difference field and the reference flow field only appears in~\eqref{eq:akeu} in the inertial production term, and for small $Re$, this is small. At early times, the quantity $\left\langle\theta_{1}^{\prime\prime}\right\rangle$ is very small, and $\left\langle\theta_{2}^{\prime\prime}\right\rangle\approx{1}$; i.e., the perturbation we impose in the conformation tensor is predominantly aligned with the direction of polymer extension. During regime (I) ($\tau\lesssim{0.1}$), the orientation changes, and $\left\langle\theta_{1}^{\prime\prime}\right\rangle$, the proportion of the uncertainty energy $\left\langle{E}_{\Delta}\right\rangle$ perpendicular to the extensional direction of the polymers, increases to approximately $0.15$. There is a further increase in $\left\langle\theta_{1}^{\prime\prime}\right\rangle$ during regime (II) ($0.1\lesssim\tau\lesssim{1}$), and during regime (III) ($1\lesssim\tau\lesssim30$), $\left\langle\theta_{1}^{\prime\prime}\right\rangle\approx0.25$. During the exponential growth regime approximately a quarter of the uncertainty energy is aligned perpendicular to the direction of polymeric extension. The increase in the proportion of the uncertainty energy aligned normal to the principal direction of reference polymer deformation prior to regime (III) is consistent with the analysis in~\S~\ref{ta}; we expect the production of uncertainty due to polymer advection $\left\langle{A}_{\Delta}\right\rangle$ to be larger when the uncertainty energy is aligned with the gradient of the polymer deformation trace, which we expect to be normal to the primary polymer deformation. In regime (IV) ($\tau\gtrsim30$), there is a slight further increase in $\left\langle\theta_{1}^{\prime\prime}\right\rangle$. That $\left\langle\theta_{1}^{\prime\prime}\right\rangle$ always remains below approximately $0.3$ is further indication that there is a limit to complete decorrelation: both realisations are subject to the same forcing, and even when their evolutions are fully diverged, the imprint of the forcing in each flow results in a residual correlation.

As a measure of the relative orientation of the uncertainty in the conformation tensor we define
\begin{equation}\psi^{\prime}_{ij}=\frac{\Delta{c}^{\prime{2}}_{ij}}{\left\langle\Pi_{\Delta}\right\rangle},\end{equation}
with an equivalent definition for $\psi^{\prime\prime}_{ij}$. \hlb{We note here that the normalisation of $\psi^{\prime}_{ij}$ by $\left\langle\Pi_{\Delta}\right\rangle$ (as opposed to $\left\langle\Gamma_{\Delta}\right\rangle$) ensures the sum over all components of $\psi_{ij}^{\prime}$ equals one.} The right panel of figure~\ref{fig:pa} shows the evolution of $\left\langle\psi^{\prime}\right\rangle$ and $\left\langle\psi^{\prime\prime}\right\rangle$.

At all times the largest component of $\left\langle\psi^{\prime}\right\rangle$ is $\left\langle\psi_{22}^{\prime}\right\rangle$, and the largest component of $\left\langle\psi^{\prime\prime}\right\rangle$ is $\left\langle\psi_{22}^{\prime\prime}\right\rangle$ - the largest component of the \hla{uncertainty in the polymer deformation} is aligned with stretching and polymeric extension in the reference flow. During regime (II) this proportion decreases, from approximately $0.6$ to approximately $0.35$ at $\tau=1$, before increasing again during regime (III) to approximately $0.5$. Following the expression for $UC1_{\Delta}$ in~\eqref{eq:uc1}, this implies that $\left\langle{UC1}_{\Delta}\right\rangle$ will be positive at all times. We next inspect the evolution of $\left\langle\psi^{\prime\prime}\right\rangle$. Note that in the right panel of figure~\ref{fig:pa} we have plotted $\left\langle\psi_{11}^{\prime\prime}\right\rangle$, $\left\langle\psi_{12}^{\prime\prime}\right\rangle$ and $\left\langle1-\psi_{22}^{\prime\prime}\right\rangle$. At short times the orientation of the \hla{uncertainty in the polymer deformation} relative to the reference polymeric deformation is roughly constant, with $\left\langle\psi_{22}^{\prime\prime}\right\rangle\approx0.997$. At the start of regime (III), from $\tau\approx{1}$ to $\tau\approx{4}$, $\left\langle\psi_{22}^{\prime\prime}\right\rangle$ \hlb{decreases}. This period coincides with exponential growth starting in $\left\langle{E}_{\Delta}\right\rangle$ whilst $\left\langle\Gamma_{\Delta}\right\rangle$ continues to decay. As we will discuss later, during this period, uncertainty is growing exponentially at large scales, but is reducing due to polymeric dissipation and relaxation at small scales. During this period at the start of regime (III), there is also a decrease in $\left\langle\psi_{11}^{\prime\prime}\right\rangle$, and an increase in $\left\langle\psi_{12}^{\prime\prime}\right\rangle$. During the transition to exponential growth, a greater proportion of the \hla{uncertainty in the polymeric deformation} is deviatoric in an orthornormal basis aligned with the reference polymeric deformation, indicating a rotation of the uncertainty relative to the polymeric deformation of the reference field as the two realisations decorrelate. However, we note that for the remainder of regime (III) ($4\lesssim\tau\lesssim30$) the orientation of \hla{uncertainty in the polymeric deformation} is roughly constant with respect to the reference polymeric deformation. At the end of regime (III), there is a sharp decrease in $\left\langle\psi_{22}^{\prime\prime}\right\rangle$ ($\left\langle1-\psi_{22}^{\prime\prime}\right\rangle$ increases by two orders of magnitude), and a corresponding increase in $\left\langle\psi_{11}^{\prime\prime}\right\rangle$ and $\left\langle\psi_{12}^{\prime\prime}\right\rangle$. \hlb{The right panel of figure~\mbox{\ref{fig:pa}} shows the evolution of $-\left(\left\langle\Gamma_{\Delta}\right\rangle-\left\langle\Pi_{\Delta}\right\rangle\right)/2\left\langle\Gamma_{\Delta}\right\rangle=-\left\langle\det\left(\Delta{c}_{ij}\right)\right\rangle/\left\langle\Gamma_{\Delta}\right\rangle$, and we note this quantity is always positive, indicating the spatial average of the determinant of $\Delta{c}_{ij}$ is negative. The relative differences between $\left\langle\Gamma_{\Delta}\right\rangle$ and $\left\langle\Pi_{\Delta}\right\rangle$ are small: $-\left(\left\langle\Gamma_{\Delta}\right\rangle-\left\langle\Pi_{\Delta}\right\rangle\right)/2\left\langle\Gamma_{\Delta}\right\rangle$ remains below $0.01$ during regimes (I) to (III), increasing to approximately $0.1$ in regime (IV). The evolution of $-\left(\left\langle\Gamma_{\Delta}\right\rangle-\left\langle\Pi_{\Delta}\right\rangle\right)/2\left\langle\Gamma_{\Delta}\right\rangle$ closely follows the evolution of $\left\langle\psi_{12}^{\prime\prime}\right\rangle$, suggesting that the exponential growth in regime (III) is characterised not only by a relative rotation of the uncertainty in polymer deformation, but also a relative increase in $\left\lvert\det\left(\Delta{c}_{ij}\right)\right\rvert$.}

\subsubsection{Terms contributing to the evolution of $\left\langle{E}_{\Delta}\right\rangle$}\label{sec:E}

\begin{figure}
\includegraphics[width=0.49\textwidth]{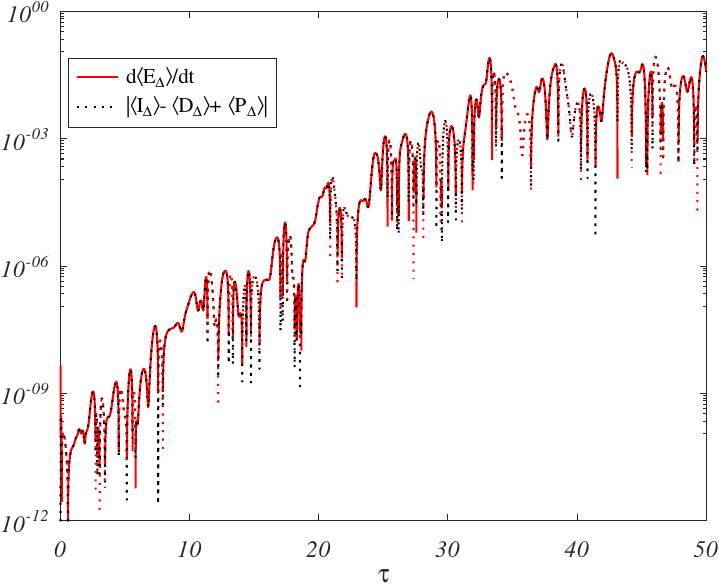}
\includegraphics[width=0.49\textwidth]{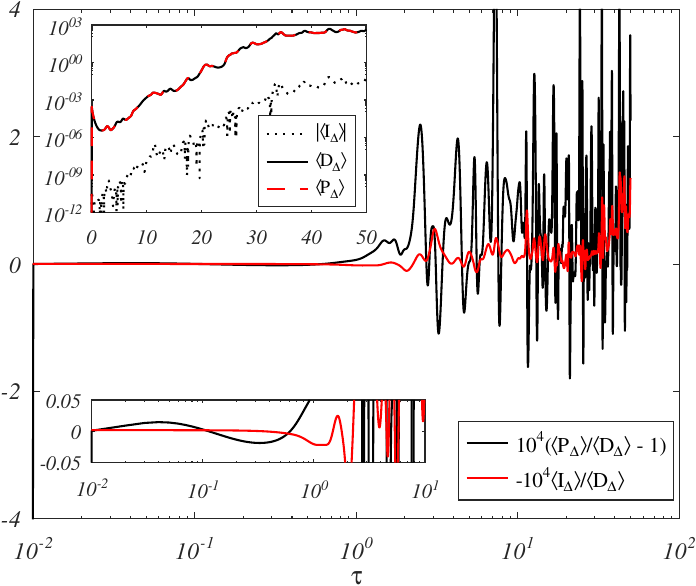}
\caption{The evolution of $d\left\langle{E}_{\Delta}\right\rangle/dt$ and terms in~\eqref{eq:akeu} for the reference configuration ($Re=10^{-2}$, $\beta=1/2$, $\varepsilon=0$, $\kappa=2.5\times{10}^{-5}$, $n=4$, $Wi=2$). The left panel shows the sum of terms in ~\eqref{eq:akeu} and the calculated values of $d\left\langle{E}_{\Delta}\right\rangle/dt$. Where $d\left\langle{E}_{\Delta}\right\rangle/dt>0$, a solid red line is used, and where $d\left\langle{E}_{\Delta}\right\rangle/dt<0$, a dotted red line. The right panel shows ratios of the terms in~\eqref{eq:akeu}. The lower inset highlights the early-time evolution. The upper inset shows the evolution of the individual terms $\left\langle{I}_{\Delta}\right\rangle$, $\left\langle{D}_{\Delta}\right\rangle$, and $\left\langle{P}_{\Delta}\right\rangle$.\label{fig:dEdtref}}
\end{figure}

We now consider how the terms in~\eqref{eq:akeu} evolve. The left panel of figure~\ref{fig:dEdtref} shows the evolution of $d\left\langle{E}_{\Delta}\right\rangle/dt$, and the terms contributing to $d\left\langle{E}_{\Delta}\right\rangle/dt$ in~\eqref{eq:akeu}. Note that we have calculated $d\left\langle{E}_{\Delta}\right\rangle/dt$ with a first order forward difference approximation based on successive values of $\left\langle{E}_{\Delta}\right\rangle$ separated by $\delta\tau=10^{-3}$. There is very close agreement between the calculated value of $d\left\langle{E}_{\Delta}\right\rangle/dt$ and $\left\langle{I}_{\Delta}\right\rangle-\left\langle{D}_{\Delta}\right\rangle+\left\langle{P}_{\Delta}\right\rangle$, providing confirmation that~\eqref{eq:akeu} holds. The right panel of figure~\ref{fig:dEdtref} shows the evolution of the ratios $-\left\langle{I}_{\Delta}\right\rangle/\left\langle{D}_{\Delta}\right\rangle$ and $\left\langle{P}_{\Delta}\right\rangle/\left\langle{D}_{\Delta}\right\rangle-1$ (scaled by $10^{4}$). The upper inset shows the individual terms $\left\langle{I}_{\Delta}\right\rangle$, $\left\langle{P}_{\Delta}\right\rangle$, and $\left\langle{D}_{\Delta}\right\rangle$. Firstly, we note that $\left\langle{I}_{\Delta}\right\rangle$ is almost always negative (we have plotted $\left\lvert\left\langle{I}_{\Delta}\right\rangle\right\rvert$ in the inset), indicating that in this regime the (small) net effect of the inertial terms is to reduce uncertainty. The quantities $\left\langle{P}_{\Delta}\right\rangle$ and $\left\langle{D}_{\Delta}\right\rangle$ are closely matched, and approximately $4$ orders of magnitude larger than $\left\langle{I}_{\Delta}\right\rangle$. The quantity $\left\langle{P}_{\Delta}\right\rangle/\left\langle{D}_{\Delta}\right\rangle-1$ provides a measure of this match: for $\left\langle{P}_{\Delta}\right\rangle/\left\langle{D}_{\Delta}\right\rangle-1>0$, polymeric \hla{propagation} of uncertainty outweighs viscous dissipation of uncertainty. For $\left\langle{P}_{\Delta}\right\rangle/\left\langle{D}_{\Delta}\right\rangle-1<0$, viscous dissipation dominates. The evolution of $\left\langle{P}_{\Delta}\right\rangle/\left\langle{D}_{\Delta}\right\rangle-1$ shows several distinct behaviours. At very early times (lower inset) the quantity $\left\langle{P}_{\Delta}\right\rangle/\left\langle{D}_{\Delta}\right\rangle-1$ is positive, indicating the early stage increase in uncertainty is driven by polymeric \hla{propagation}. Subsequently, $\left\langle{P}_{\Delta}\right\rangle/\left\langle{D}_{\Delta}\right\rangle-1$ becomes negative on timescales of the order of $10^{-1}$. In this period, there is a net reduction in uncertainty as viscous dissipation dominates polymeric \hla{propagation}. From $\tau\approx{1}$ to $\tau\approx{25}$, $\left\langle{P}_{\Delta}\right\rangle/\left\langle{D}_{\Delta}\right\rangle-1$ fluctuates, but is predominantly positive. This period corresponds to the regime of exponential growth of uncertainty. As uncertainty saturates, the quantity $\left\langle{P}_{\Delta}\right\rangle/\left\langle{D}_{\Delta}\right\rangle-1$ increases, with a trend matching that of $-\left\langle{I}_{\Delta}\right\rangle/\left\langle{D}_{\Delta}\right\rangle$. 

\subsubsection{Terms contributing to the evolution of $\left\langle{\Gamma}_{\Delta}\right\rangle$}\label{sec:G}

\begin{figure}
\includegraphics[width=0.49\textwidth]{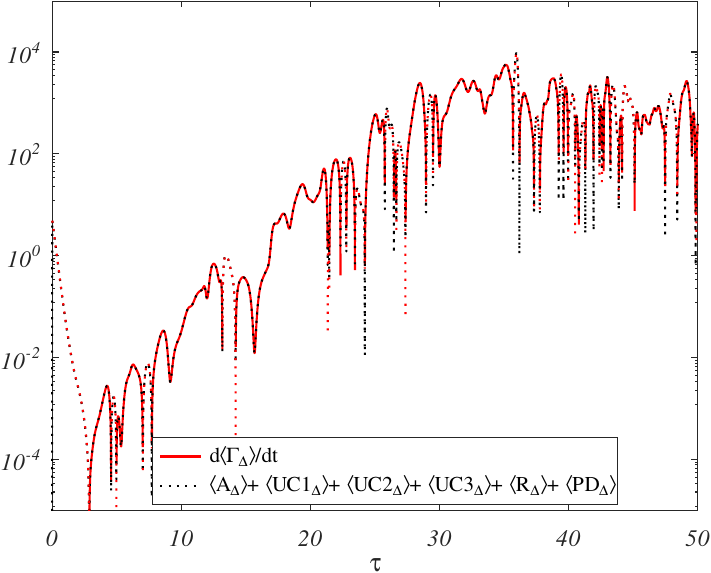}
\includegraphics[width=0.49\textwidth]{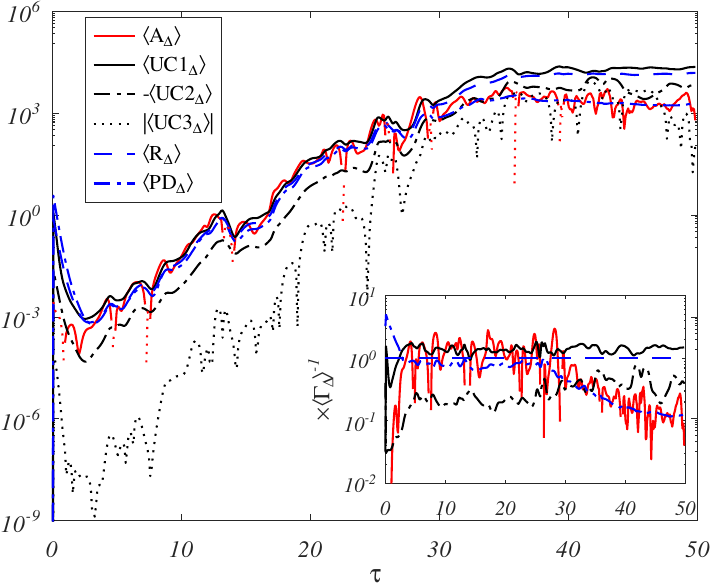}
\caption{The evolution of $d\left\langle{\Gamma}_{\Delta}\right\rangle/dt$ and terms in~\eqref{eq:dG} for the reference configuration ($Re=10^{-2}$, $\beta=1/2$, $\varepsilon=0$, $\kappa=2.5\times{10}^{-5}$, $n=4$, $Wi=2$). The left panel shows the sum of terms in ~\eqref{eq:dG} and the calculated values of $d\left\langle{\Gamma}_{\Delta}\right\rangle/dt$. Where $d\left\langle{\Gamma}_{\Delta}\right\rangle/dt>0$, a solid red line is used, and where $d\left\langle{\Gamma}_{\Delta}\right\rangle/dt<0$, a dotted red line. The right panel shows the individual terms in~\eqref{eq:dG}. The inset shows these same terms normalised by $\left\langle\Gamma_{\Delta}\right\rangle$. Where $\left\langle{A}_{\Delta}\right\rangle$ is negative, it is plotted with a dotted red line, and with a solid red line where positive.\label{fig:dGdtref}}
\end{figure}

Figure~\ref{fig:dGdtref} shows the evolution of $d\left\langle{\Gamma}_{\Delta}\right\rangle/dt$, and the contributing terms in~\eqref{eq:dG}. The left panel shows the evolution of $d\left\langle{\Gamma}_{\Delta}\right\rangle/dt$ (evaluated from successive values of $\left\langle\Gamma_{\Delta}\right\rangle$ at intervals of $\delta{\tau}=10^{-3}$), and the sum of the terms in the right hand side of~\eqref{eq:dG}. The two quantities match very closely, with discrepancies only arising when $d\left\langle{\Gamma}_{\Delta}\right\rangle/dt$ is rapidly changing, due to the first order finite difference approximation used to evaluate $d\left\langle{\Gamma}_{\Delta}\right\rangle/dt$.

The right panel of figure~\ref{fig:dGdtref} shows the evolution of the individual terms in~\eqref{eq:dG}. For this configuration with $\varepsilon=0$ (i.e. Oldroyd B), $\left\langle{R}_{\Delta}\right\rangle=\left\langle2\Gamma_{\Delta}/Wi\right\rangle=\left\langle\Gamma_{\Delta}\right\rangle$. We first consider the upper convected terms. $\left\langle{UC1}_{\Delta}\right\rangle$ is at all times positive. This observation is consistent with the finding in~\S~\ref{sec:orient} that $\left\langle\psi_{22}^{\prime}\right\rangle$ is always the largest component of $\left\langle\psi^{\prime}\right\rangle$. Stretching in the conformation tensor difference field is predominantly aligned with stretching in the reference flow, and this \hla{results in a production of} uncertainty. $\left\langle{UC2}_{\Delta}\right\rangle$ is at all times negative (note we have plotted $-\left\langle{UC2}_{\Delta}\right\rangle$), and always approximately an order of magnitude smaller than $\left\langle{UC1}_{\Delta}\right\rangle$. $\left\langle{UC3}_{\Delta}\right\rangle$ is sometimes positive and sometimes negative, and with magnitude several orders smaller than $\left\langle{UC1}_{\Delta}\right\rangle$, as predicted in~\S~\ref{ta}. The relative importance of $\left\langle{UC3}_{\Delta}\right\rangle$ increases with time as the uncertainty increases, but even in regime (IV) it remains an order of magnitude smaller than the other two upper convected terms. Of the upper convected terms, it is $\left\langle{UC1}_{\Delta}\right\rangle$ which dominates the dynamics of uncertainty evolution, with $\left\langle{UC2}_{\Delta}\right\rangle\sim-0.1\left\langle{UC2}_{\Delta}\right\rangle$ through regime (III), and $\left\langle{UC1}_{\Delta}\right\rangle>0$ at all times. The net contribution of the upper convected terms is to amplify uncertainty.

The inset of the right panel of figure~\ref{fig:dGdtref} shows the evolution of the terms in~\eqref{eq:dG} normalised by $\left\langle\Gamma_{\Delta}\right\rangle$. At early times, $\left\langle{PD}_{\Delta}\right\rangle$ dominates, as the perturbation is focused at small scales. Over the first few time units, the relative magnitude of $\left\langle{PD}_{\Delta}\right\rangle$ reduces, and during the exponential regime ($\tau\in\left[1,30\right]$), we see $\left\langle{PD}_{\Delta}\right\rangle\approx\left\langle{R}_{\Delta}\right\rangle=\left\langle\Gamma_{\Delta}\right\rangle$. During the period of exponential growth of $\left\langle\Gamma_{\Delta}\right\rangle$, the ratio $\left\langle{UC1}_{\Delta}\right\rangle/\left\langle\Gamma_{\Delta}\right\rangle$ is always greater than unity, and the ratio $\left\langle{PD}_{\Delta}\right\rangle/\left\langle\Gamma_{\Delta}\right\rangle$ is always less than unity. The growth of uncertainty is approximately determined by the balance of the quantity 
\begin{equation}\frac{\left\langle{UC1}_{\Delta}\right\rangle+\left\langle{A}_{\Delta}\right\rangle-\left\langle{R}_{\Delta}\right\rangle-\left\langle{PD}_{\Delta}\right\rangle}{\left\langle\Gamma_{\Delta}\right\rangle},\end{equation}
which is equivalent (neglecting the contributions of $\left\langle{UC2}_{\Delta}\right\rangle$ and $\left\langle{UC3}_{\Delta}\right\rangle$), to twice the maximal Lyapunov exponent of $\left\langle\Gamma_{\Delta}\right\rangle$. With $\left\langle{R}_{\Delta}\right\rangle=\left\langle{\Gamma}_{\Delta}\right\rangle$, and $\left\langle{UC1}_{\Delta}\right\rangle$ and $\left\langle{PD}_{\Delta}\right\rangle$ closely correlated to $\left\langle{\Gamma}_{\Delta}\right\rangle$, it is the advective term which largely determines whether uncertainty increases or reduces (note that this advective term, originating from~\eqref{eq:cte}, remains relevant even at very low $Re$). This can be seen in figure~\ref{fig:dGdtref}, where regions of negative $\left\langle{A}_{\Delta}\right\rangle$ (shown with a dotted red line), correspond to short term decreases in $\left\langle{R}_{\Delta}\right\rangle$. 

\subsubsection{The spectra of uncertainty}

\begin{figure}
\includegraphics[width=0.49\textwidth]{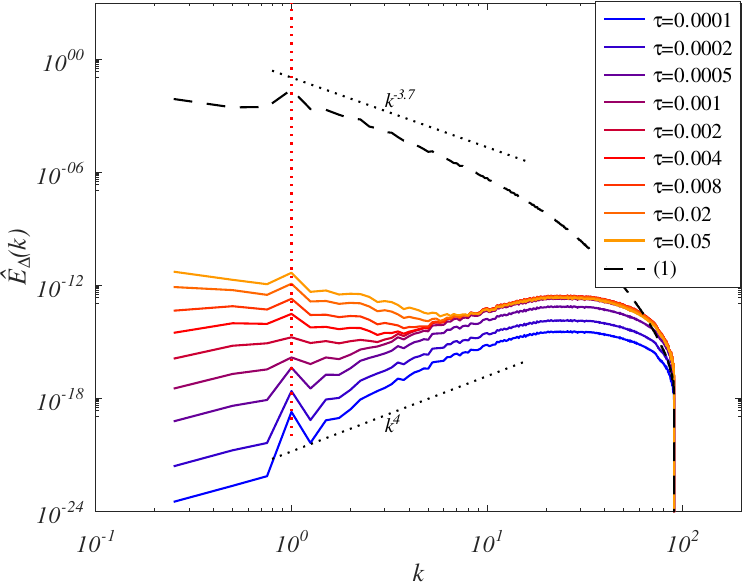}
\includegraphics[width=0.49\textwidth]{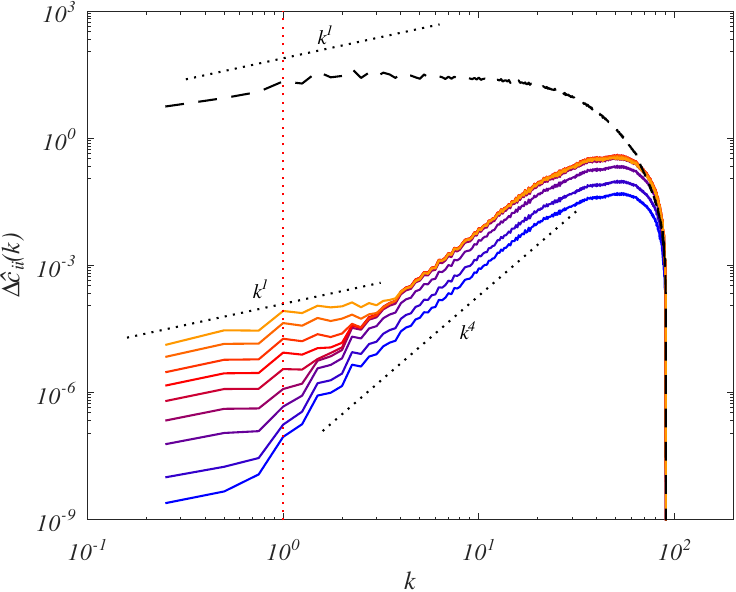}
\caption{Evolution of the uncertainty energy spectrum $\hat{E}_{\Delta}\left(k\right)$ (left panel), and the uncertainty in the conformation tensor trace ${\Delta\hat{c}}_{ii}\left(k\right)$ (right panel) for the reference configuration ($Re=10^{-2}$, $\beta=1/2$, $\varepsilon=0$, $\kappa=2.5\times{10}^{-5}$, $n=4$, $Wi=2$) at very short times (regime (I)). The dashed black lines show the reference energy spectrum $\hat{E}^{\left(1\right)}$ and conformation tensor trace $\hat{c}^{\left(1\right)}\left(k\right)$. The dotted red line indicates the forcing wavenumber $k=2\pi$.\label{fig:dspec_evol_veryshort}}
\end{figure}

\begin{figure}
\includegraphics[width=0.49\textwidth]{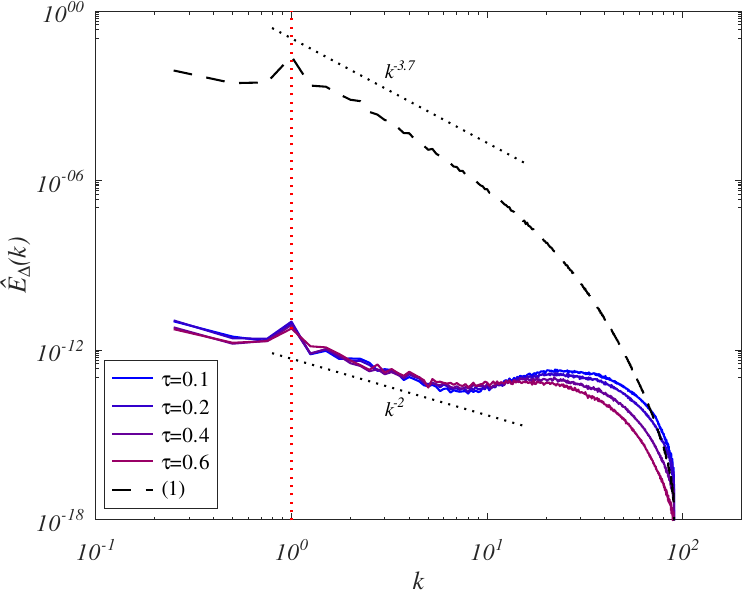}
\includegraphics[width=0.49\textwidth]{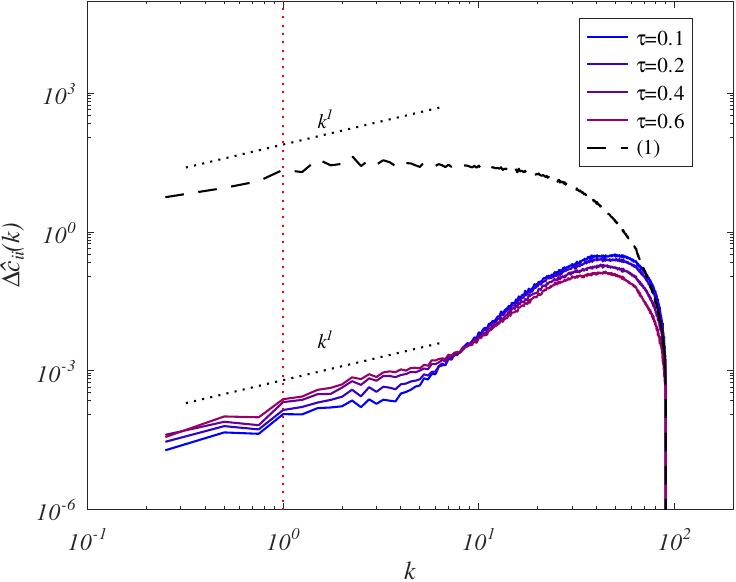}
\caption{Evolution of the uncertainty energy spectrum $\hat{E}_{\Delta}\left(k\right)$ (left panel), and the uncertainty in the conformation tensor trace ${\Delta\hat{c}}_{ii}\left(k\right)$ (right panel) for the reference configuration ($Re=10^{-2}$, $\beta=1/2$, $\varepsilon=0$, $\kappa=2.5\times{10}^{-5}$, $n=4$, $Wi=2$) at short times (regime (II)). The dashed black lines show the reference energy spectrum $\hat{E}^{\left(1\right)}$ and conformation tensor trace $\hat{c}^{\left(1\right)}\left(k\right)$. The dotted red line indicates the forcing wavenumber $k=2\pi$.\label{fig:dspec_evol_short}}
\end{figure}

\begin{figure}
\includegraphics[width=0.49\textwidth]{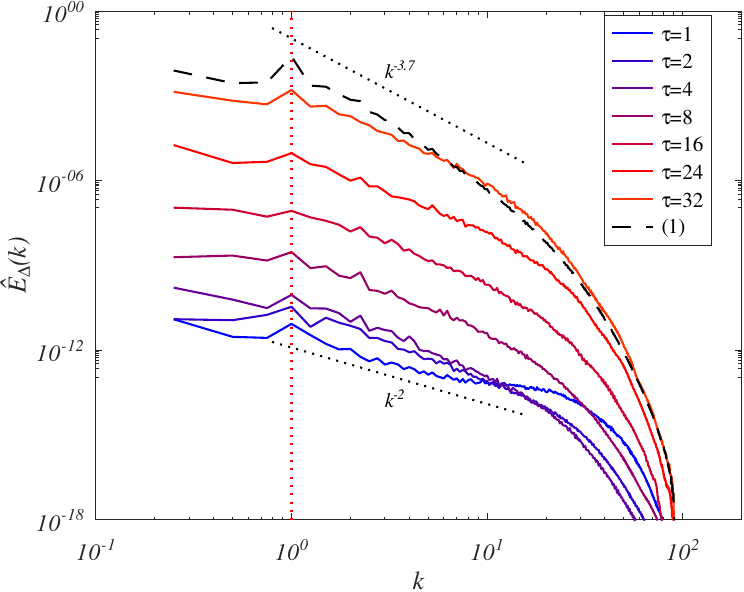}
\includegraphics[width=0.49\textwidth]{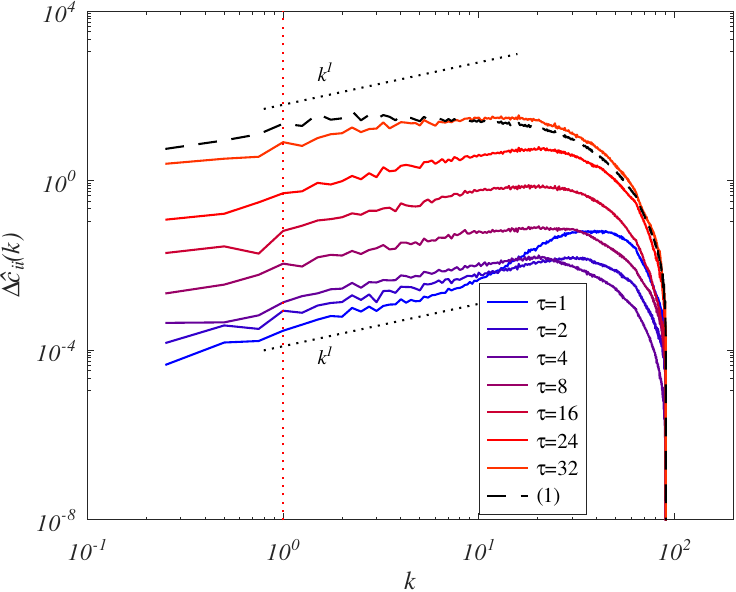}
\caption{The evolution of the uncertainty energy spectrum $\hat{E}_{\Delta}\left(k\right)$ (left panel), and the uncertainty in the conformation tensor trace ${\Delta\hat{c}}_{ii}\left(k\right)$ (right panel) for the reference configuration ($Re=10^{-2}$, $\beta=1/2$, $\varepsilon=0$, $\kappa=2.5\times{10}^{-5}$, $n=4$, $Wi=2$) at longer times (regime (III)). The dashed black lines show the reference energy spectrum $\hat{E}^{\left(1\right)}$ and conformation tensor trace $\hat{c}^{\left(1\right)}\left(k\right)$. The dotted red line indicates the forcing wavenumber $k=2\pi$.\label{fig:dspec_evol}}
\end{figure}

We next investigate how the spectra of the uncertainty evolves. We calculate the energy spectra of $\Delta{u}_{i}$ (denoted $\hat{E}_{\Delta}$), and the spectra of $\Delta{c}_{ii}$ (denoted $\Delta\hat{}{c}_{ii}$). Figure~\ref{fig:dspec_evol_veryshort} shows $\hat{E}_{\Delta}$ (left panel) and ${\Delta\hat{}{c}}_{ii}$ (right panel) at very early times $\tau\le0.05$, covering until nearly the end of regime (I). Figure~\ref{fig:dspec_evol_short} shows the same for $\tau\in\left[0.1,0.6\right]$, covering regime (II). Figure~\ref{fig:dspec_evol} shows the same at later times $\tau\in\left[1,32\right]$, covering regime (III). In all three figures, the forcing wavenumber is indicated by a vertical dotted red line, and the spectra of the reference field (averaged over $50$ time units) is shown with a black dashed line.

At early times (figure~\ref{fig:dspec_evol_veryshort}) the uncertainty energy and uncertainty in the conformation tensor trace are predominantly at high wavenumbers. This is due to the nature of the perturbation, which is imposed via~\eqref{eq:ptbn}. At the earliest times, both $\hat{}{E}_{\Delta}$ and ${\Delta\hat{c}}_{ii}$ have a slope of $k^{4}$, consistent with the $\nabla^{4}$ operator used to impose the perturbation. For $\tau\le10^{-3}$, there is an increase in $\hat{E}_{\Delta}$ and ${\Delta\hat{c}}_{ii}$ across all wavenumbers, as the perturbation is applied over a finite time - the characteristic time-scale of the perturbation in~\eqref{eq:ptbn} is $10^{-3}$. From the earliest times, there is an almost immediate increase in uncertainty at large scales, and this increase continues after the perturbation has ceased, and spreads across a greater range of wavenumbers with increasing $\tau$. This transfer of uncertainty from small scales to large scales is due to the elliptic constraint of incompressibility: small scale uncertainty localised at one point in space creates uncertainty everywhere, instantly. 

At slightly longer times - $\tau\in\left[0.1,1\right]$ (figure~\ref{fig:dspec_evol_short}), there is a reduction in uncertainty at small scales, and the uncertainty at large scales remains roughly constant. This period corresponds to regime (II) as defined above, and the time during which the viscous dissipation of uncertainty dominates polymeric \hla{propagation}. We see a slight reduction in $\hat{E}_{\Delta}$ across all scales, and a decrease in ${\Delta\hat{c}}_{ii}$ at small scales. In this regime the spectra  $\hat{E}_{\Delta}$ has a slope of approximately $k^{-2}$, and ${\Delta\hat{c}}_{ii}$ has a slope (which matches that of $\hat{{c}}^{\left(1\right)}_{ii}$) of $k^{1}$. After this (regime (III), figure~\ref{fig:dspec_evol}), there is exponential growth of $\hat{E}_{\Delta}$ and ${\Delta\hat{c}}_{ii}$. This exponential growth occurs initially at large scales, whilst diffusion still dominates small scales until $\tau\approx4$, after which we see exponential growth across all scales. With increasing $\tau$, the slope of $\hat{E}_{\Delta}$ increases towards $k^{-3.7}$, the slope of the reference energy spectra.

\begin{figure}
\includegraphics[width=0.49\textwidth]{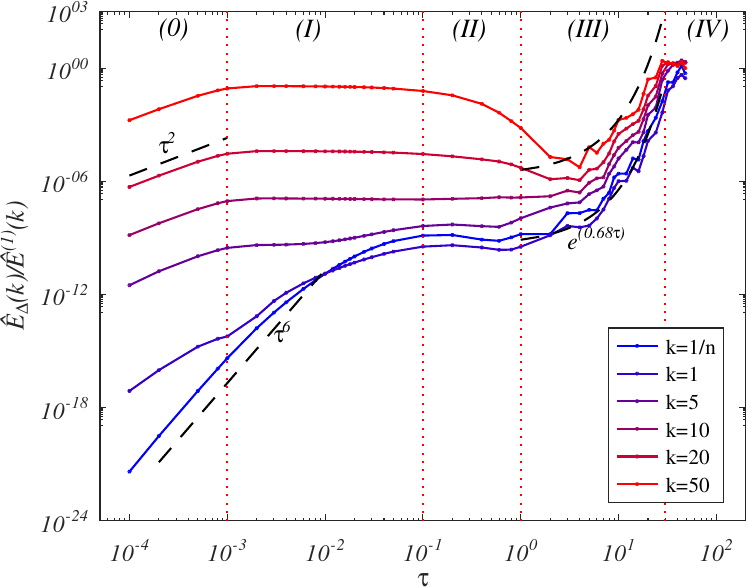}
\includegraphics[width=0.49\textwidth]{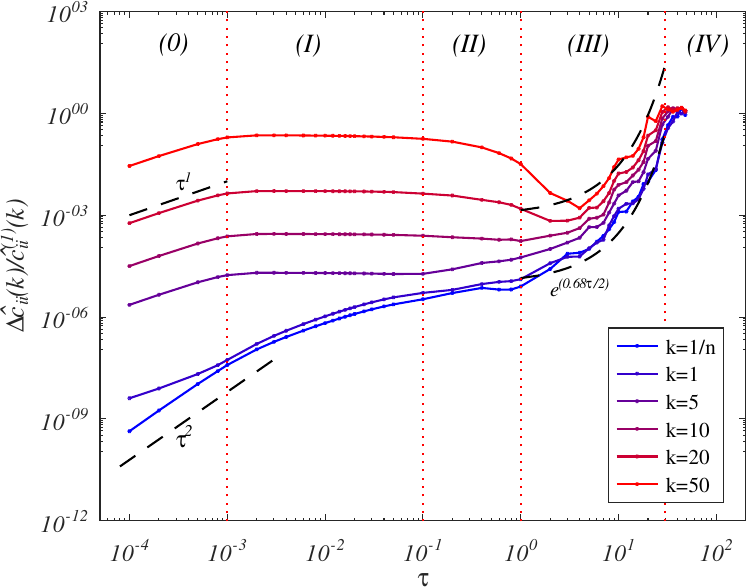}
\caption{The time evolution of components of the uncertainty energy spectra $\hat{E}_{\Delta}\left(k\right)$ (left panel), and the spectra of the uncertainty in conformation tensor trace ${\Delta\hat{c}}_{ii}\left(k\right)$ (right panel), normalised by the reference spectra, for the reference configuration ($Re=10^{-2}$, $\beta=1/2$, $\varepsilon=0$, $Sc=4\times{10}^{6}$, $n=4$, $Wi=2$).\label{fig:dspec_v_time}}
\end{figure}

In figure~\ref{fig:dspec_v_time} we plot the time evolution of individual components of $\hat{E}_{\Delta}$ (left panel) and ${\Delta\hat{c}}_{ii}$ (right panel). The different regimes of uncertainty evolution are marked with vertical dotted red lines. We denote regime (0) as the timescale over which the perturbation is imposed. For $\hat{E}_{\Delta}$, the uncertainty at small scales initially grows at a rate proportional to $\tau^{2}$, which is consistent with the injection of uncertainty during the imposition of the perturbation for $\tau<\delta_{t0}$. For $\tau>\delta_{t0}$, this growth at small scales ceases. For ${\Delta\hat{c}}_{ii}$, this early growth at small scales is at a rate proportional to $\tau$. The picture is different at large scales, with $\hat{E}_{\Delta}$ growing proportional to $\tau^{6}$, and ${\Delta\hat{c}}_{ii}$ growing proportional to $\tau^{2}$. This growth at large scales continues for $\tau>\delta_{t0}$. We have also performed simulations with $\delta_{t0}=10^{-5}\ll\delta{t}=4\times10^{-4}$, where $\delta{t}$ is the computational time step.  Even in this limit, the same short-time growth rate of $\tau^{6}$ is observed, and this growth rate persists for the same duration (i.e. it persists until $\tau\gg\delta_{t0}$). 

Through regime (I), the uncertainty in both flow and conformation tensor continues to grow at large scales, whilst the uncertainty at small scales is roughly constant. The growth rate decreases with time, as the balance between viscous dissipation and polymeric \hla{propagation} changes. By $\tau\approx0.1$, viscous dissipation of uncertainty becomes greater than polymeric \hla{propagation}, and the uncertainty in both flow and conformation tensor decreases at small scales through regime (II). For $\tau>1$, in regime (III), there is exponential growth of uncertainty in both the flow and conformation tensor, again initially at large scales, and followed by small scales by $\tau\approx{4}$. For $\tau>30$, in regime (IV), the uncertainty saturates at all scales.

We comment here on the short time evolution of uncertainty. If we impose a perturbation on $c_{ij}^{\left(2\right)}$ using a Laplacian operator (rather than a hyperviscous operator), we observe qualitatively similar behaviour during regimes (I) and (II). We note that for all perturbations, regimes (III) and (IV) are unchanged; the growth rate in regime (III), and the saturation values in regime (IV), are independent of the perturbation imposed. 

\subsection{The effect polymeric diffusivity $\kappa$}\label{sec:sc_nl}

\begin{figure}
\includegraphics[width=0.49\textwidth]{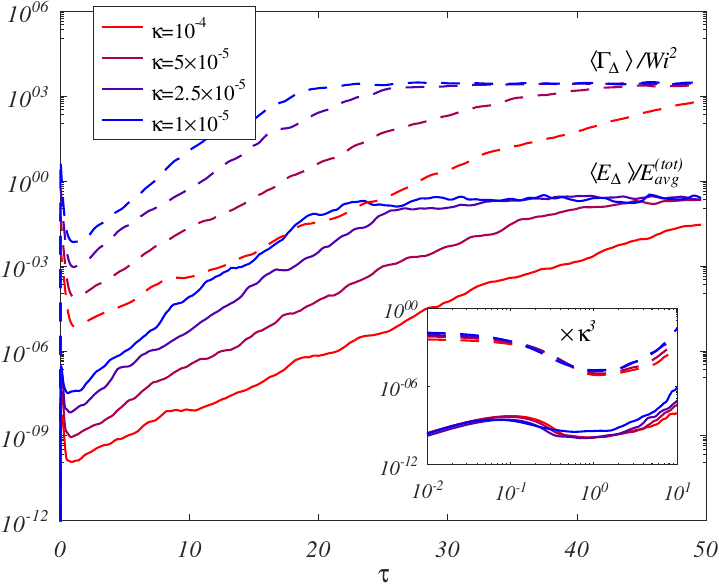}
\includegraphics[width=0.49\textwidth]{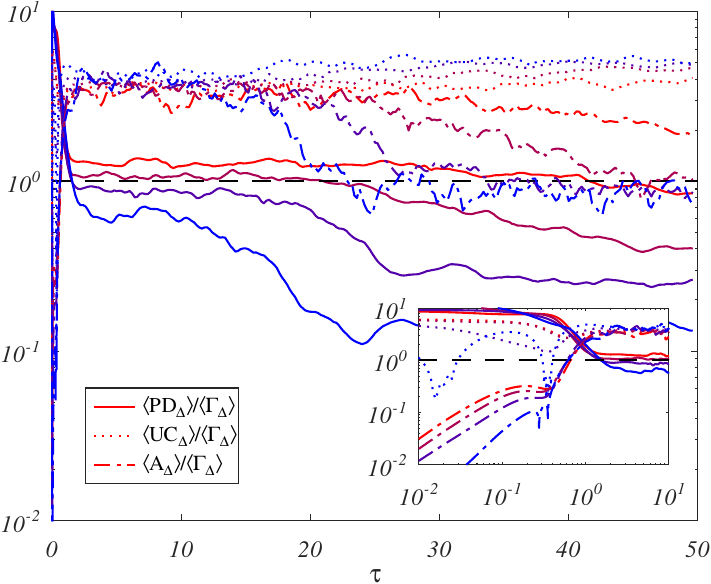}
\caption{Left panel: The evolution of $\left\langle{E}_{\Delta}\right\rangle/E_{avg}^{\left(tot\right)}$ and $\left\langle{\Gamma}_{\Delta}\right\rangle/Wi^{2}$ for different values of $\kappa$. Right panel: the evolution of terms in~\eqref{eq:dG}, normalised by $\left\langle\Gamma_{\Delta}\right\rangle$ for different values of $\kappa$. All other parameters match the reference configuration ($Re=10^{-2}$, $\beta=1/2$, $\varepsilon=0$, $n=4$, $Wi=2$). \hlb{For $\kappa=10^{-5}$, the resolution is $\left(192n\right)^{2}$, for all larger $\kappa$, the resolution is $\left(128n\right)^{2}$.}\label{fig:dEG_varsc}}
\end{figure}

Polymeric diffusivity acts to limit the smallest length scales of the polymer deformation field, and we see from~\eqref{eq:dG} that the effect of polymeric diffusivity on uncertainty is to reduce it. The left panel of figure~\ref{fig:dEG_varsc} shows the evolution of $\left\langle{E}_{\Delta}\right\rangle/E_{avg}^{\left(tot\right)}$ and $\left\langle{\Gamma}_{\Delta}\right\rangle/Wi^{2}$ for \hlb{$\kappa\in\left[10^{-4},5\times{10}^{-5},2.5\times{10}^{-5},10^{-5}\right]$}. The growth rate of uncertainty during regime (III) increases with decreasing $\kappa$. At short times, the polymeric diffusivity has a significant influence on the uncertainty. The inset shows the short time evolution of $\left\langle{E}_{\Delta}\right\rangle/E_{avg}^{\left(tot\right)}$ and $\left\langle{\Gamma}_{\Delta}\right\rangle/Wi^{2}$, collapsing when scaled by $\kappa^{3}$. This dependence on $\kappa$ is a result of the form of the perturbation: the perturbation imposed generates polymeric deformation difference spectra which scales with $k^{4}$ (as in figure~\ref{fig:dspec_evol_veryshort}). Polymeric diffusivity limits the smallest length scales of the polymer deformation field, and consequently the imposed perturbation is smaller for larger $\kappa$. The right panel of figure~\ref{fig:dEG_varsc} shows the evolution of the terms in~\eqref{eq:dG}, normalised by $\left\langle\Gamma_{\Delta}\right\rangle$, for different $\kappa$. The inset shows the early time evolution. \hlb{For $\kappa\ge2.5\times{10}^{-5}$, t}he relative magnitude of $\left\langle{UC1}_{\Delta}\right\rangle$ and $\left\langle{A}_{\Delta}\right\rangle$ is little changed by $\kappa$. At very short times, $\left\langle{PD}_{\Delta}\right\rangle/\left\langle\Gamma_{\Delta}\right\rangle$ is independent of $\kappa$. During regime (III), $\left\langle{PD}_{\Delta}\right\rangle/\left\langle\Gamma_{\Delta}\right\rangle$ is approximately constant, with a value \hlb{which decreases with decreasing} $\kappa$. \hlb{One could extrapolate this trend to estimate the maximum permissible value of $\kappa$ such that the effect of polymeric diffusivity on the chaotic dynamics is negligible (e.g. the value of $\kappa$ such that $\left\langle{PD}_{\Delta}\right\rangle/\left\langle\Gamma_{\Delta}\right\rangle<0.1$ during regime (III)), but in the absence of extremely high-fidelity simulations at very small $\kappa$ to confirm such an extrapolation, we leave this as a suggestion for future studies.} For all \hlb{four} values of $\kappa$, the terms $\left\langle{UC1}_{\Delta}\right\rangle$ and $\left\langle{A}_{\Delta}\right\rangle$ are larger than $\left\langle{PD}_{\Delta}\right\rangle$ during regime (III).

\subsection{The influence of $Re$}

\begin{figure}
\includegraphics[width=0.49\textwidth]{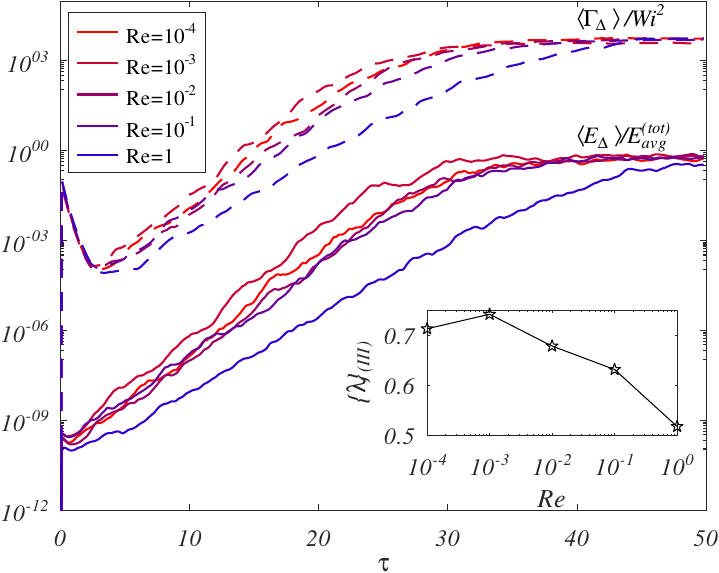}
\includegraphics[width=0.49\textwidth]{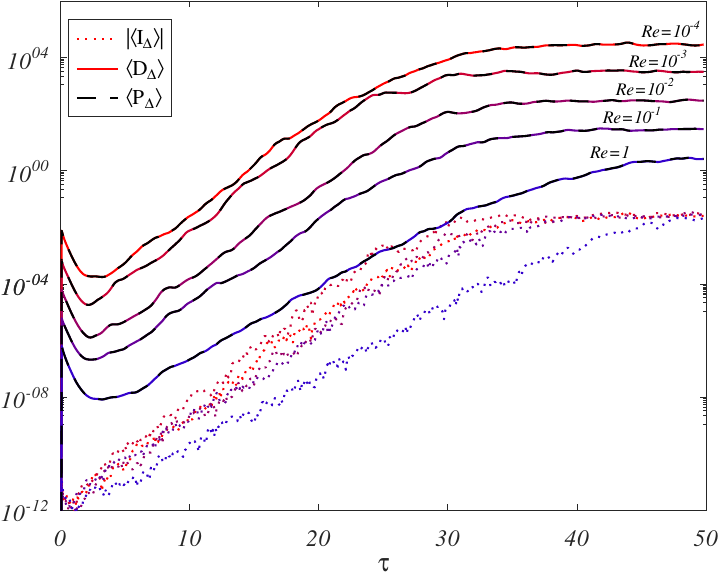}
\caption{Left panel: The evolution of $\left\langle{E}_{\Delta}\right\rangle/E_{avg}^{\left(tot\right)}$ and $\left\langle\Gamma_{\Delta}\right\rangle/Wi^{2}$ for a range of $Re$. All other parameters match the reference configuration ($\beta=1/2$, $\varepsilon=0$, $\kappa=2.5\times{10}^{-5}$, $n=4$, $Wi=2$). Inset shows the average growth rate over regime (III). Right panel: The evolution of $\left\langle{I}_{\Delta}\right\rangle$, $\left\langle{D}_{\Delta}\right\rangle$, and $\left\langle{P}_{\Delta}\right\rangle$ for a range of $Re$.\label{fig:dEG_varRe}}
\end{figure}

We next investigate the influence of $Re$, keeping all other parameters fixed. The left panel of figure~\ref{fig:dEG_varRe} shows the evolution of $\left\langle{E}_{\Delta}\right\rangle/E_{avg}^{\left(tot\right)}$ and $\left\langle\Gamma_{\Delta}\right\rangle/Wi^{2}$ for a range $Re\in\left[10^{-4},10^{-3},10^{-2},10^{-1},1\right]$. The growth rate of uncertainty during regime (III) is shown in the inset, and appears weakly dependent on $Re$ for $Re<0.1$. For $Re={1}$, the growth rate decreases. The right panel of figure~\ref{fig:dEG_varRe} shows the evolution of $\left\langle{I}_{\Delta}\right\rangle$, $\left\langle{D}_{\Delta}\right\rangle$, and $\left\langle{P}_{\Delta}\right\rangle$. The evolution of $\left\lvert\left\langle{I}_{\Delta}\right\rangle\right\rvert$ grows in approximately proportion to $\left\langle{E}_{\Delta}\right\rangle$, with the proportionality independent of $Re$. The magnitudes of $\left\langle{D}_{\Delta}\right\rangle$, and $\left\langle{P}_{\Delta}\right\rangle$ scale with $Re^{-1}$. The quantities $\left\langle{P}_{\Delta}\right\rangle/\left\langle{D}_{\Delta}\right\rangle-1$ and $-\left\langle{I}_{\Delta}\right\rangle/\left\langle{D}_{\Delta}\right\rangle$ also scale with $Re^{-1}$; this collapse is anticipated provided the average flow is independent of $Re$, as both $\left\langle{P}_{\Delta}\right\rangle$ and $\left\langle{D}_{\Delta}\right\rangle$ contain $Re$ in the denominator. What is remarkable is that the average ratio of these quantities during regime (III) is only weakly dependent on $Re$. Defining
\begin{equation}C\left(Re\right)=\frac{\left\{\left\langle{I}_{\Delta}\right\rangle/\left\langle{D}_{\Delta}\right\rangle\right\}_{\left(III\right)}}{\left[\left\{\left\langle{P}_{\Delta}\right\rangle/\left\langle{D}_{\Delta}\right\rangle\right\}_{\left(III\right)}-1\right]},\end{equation}
in which $C\left(Re\right)$ is a function of $Re$, we find \hla{$C\in\left[0.113,0.117,0.121,0.143,0.153\right]$} for $Re\in\left[10^{-4},10^{-3},10^{-2},10^{-1},1\right]$. The relative importance of the inertial production of uncertainty is still of the order of $10\%$ even for $Re=10^{-4}$. How $C\left(Re\right)$ approaches zero in the limit $Re\to0$ is an interesting question, but one which we leave for future studies. We have also calculated  the ratios $\left\{\left\langle{A}_{\Delta}\right\rangle/\left\langle{R}_{\Delta}\right\rangle\right\}_{\left(III\right)}$ and $\left\{\left\langle{UC1}_{\Delta}\right\rangle/\left\langle{R}_{\Delta}\right\rangle\right\}_{\left(III\right)}$ and find that both ratios are independent of $Re$. 

\begin{figure}
\includegraphics[width=0.99\textwidth]{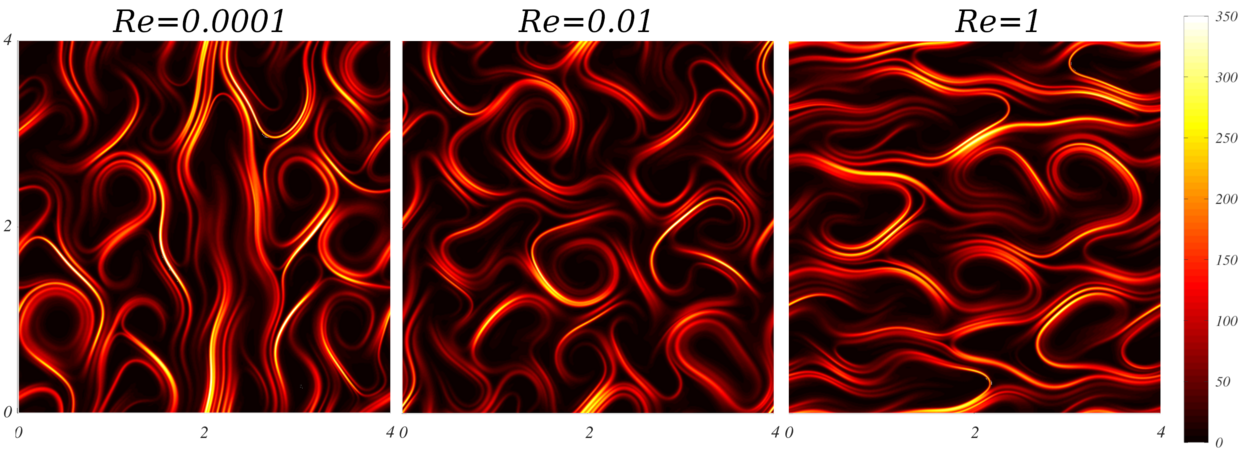}
\caption{Snapshots of the conformation tensor trace field for $Re\in\left[10^{-4},10^{-2},1\right]$. All other parameters match the reference configuration ($\beta=1/2$, $\varepsilon=0$, $\kappa=2.5\times{10}^{-5}$, $n=4$, $Wi=2$).\label{fig:trcRe}}
\end{figure}

Figure~\ref{fig:trcRe} shows snapshots of the conformation tensor trace field for three values of $Re$. Note that in these snapshots, structures similar to the ``narwhal'' or ``arrowhead'' structures recently identified as exact travelling wave solutions~\citep{page_2020,morozov_2022} are visible for both $Re=10^{-4}$ and $Re=1$. These larger scale structures spanning the domain and breaking the cellular structure of the flow are present \emph{intermittently} at all $Re$ investigated, but are present for a greater proportion of the time at larger $Re$. We also note that the occurrence of these structures is less frequent at larger $\varepsilon$: we do not observe them at all for simulations with $\varepsilon=10^{-2}$. 

\begin{figure}
\includegraphics[width=0.49\textwidth]{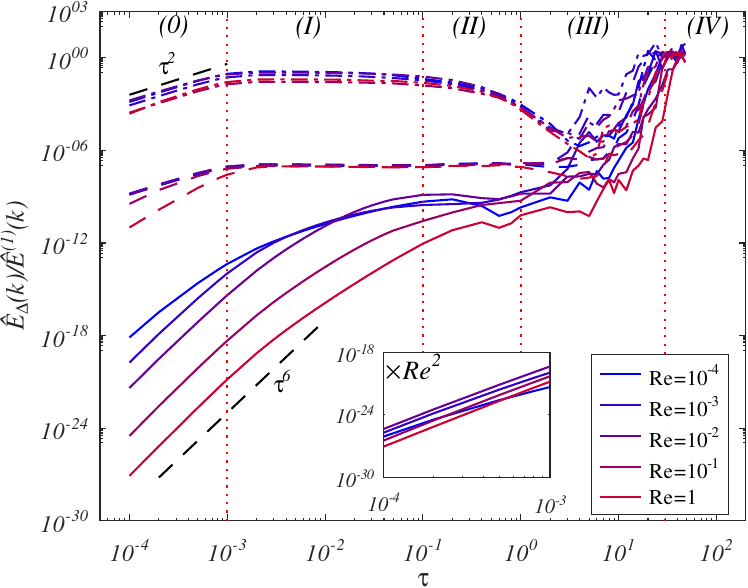}
\includegraphics[width=0.49\textwidth]{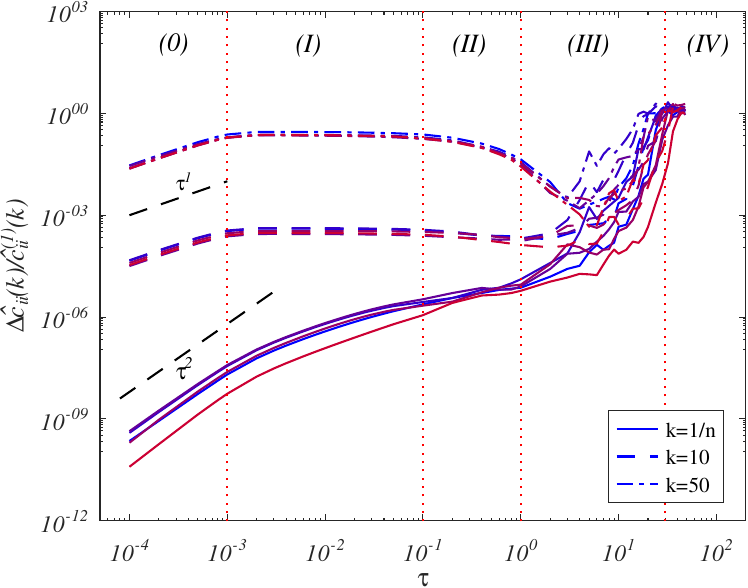}
\caption{The time evolution of components of the uncertainty energy spectra $\hat{E}_{\Delta}\left(k\right)$ (left panel), and the spectra of the uncertainty in conformation tensor trace ${\Delta\hat{c}}_{ii}\left(k\right)$ (right panel), normalised by the reference spectra, for different values of $Re$. The inset of the left panel shows components of $\hat{E}_{\Delta}\left(k\right)$ scaled by $Re^{2}$. The line styles correspond to different wavenumbers, with $k=1/n$ - solid lines, $k=10$ - dashed lines, and $k=50$ - dash-dot lines. All other parameters match the reference configuration ($\beta=1/2$, $\varepsilon=0$, $\kappa=2.5\times{10}^{-5}$, $n=4$, $Wi=2$).\label{fig:dspec_v_time_varRe}}
\end{figure}

We next assess the influence of $Re$ on the spectra of uncertainty, following the same analysis as above for the reference configuration. The results are shown in figure~\ref{fig:dspec_v_time_varRe}. From figure~\ref{fig:dEG_varRe}, we have observed the evolution of uncertainty in regime (III) is only weakly dependent on $Re$, and here we are interested in how the inertial terms influence the early stages of uncertainty growth. At short times, for all $Re\in\left[10^{-4},1\right]$ we see the same growth rates: at the largest scales, $\hat{E}_{\Delta}$ grows with $\tau^{6}$, and this growth rate persists for some time after the imposition of the perturbation. At the smallest scales, $\hat{E}_{\Delta}$ grows with $\tau^{2}$ during the imposition of the perturbation, after which it plateaus, before diffusive effects begin to reduce uncertainty. Although the growth rate is independent of $Re$, the magnitude of $\hat{E}_{\Delta}$ at small $k$ scales approximately with $Re^{-2}$. This can be seen in the inset of the left panel of figure~\ref{fig:dspec_v_time_varRe}, which shows the early stages of the evolution of $Re^{2}\hat{E}_{\Delta}$. There appears to be a maximum uncertainty in the large scales during regime (II). For smaller $Re$, the growth proportional to $\tau^{6}$ persists for a shorter time. For all $Re\le{1}$, the low wavenumber components of $\hat{E}_{\Delta}$ reach a similar magnitude during the transition from regime (I) to regime (II), after which the evolution during the exponential growth regime (III) tracks approximately the same course. For $Re=1$, the evolution of the large scales of uncertainty is qualitatively similar, but with smaller magnitudes and lower growth rates during the exponential regime (III).

\subsection{The effect of increasing $Wi$}

\begin{figure}
\includegraphics[width=0.99\textwidth]{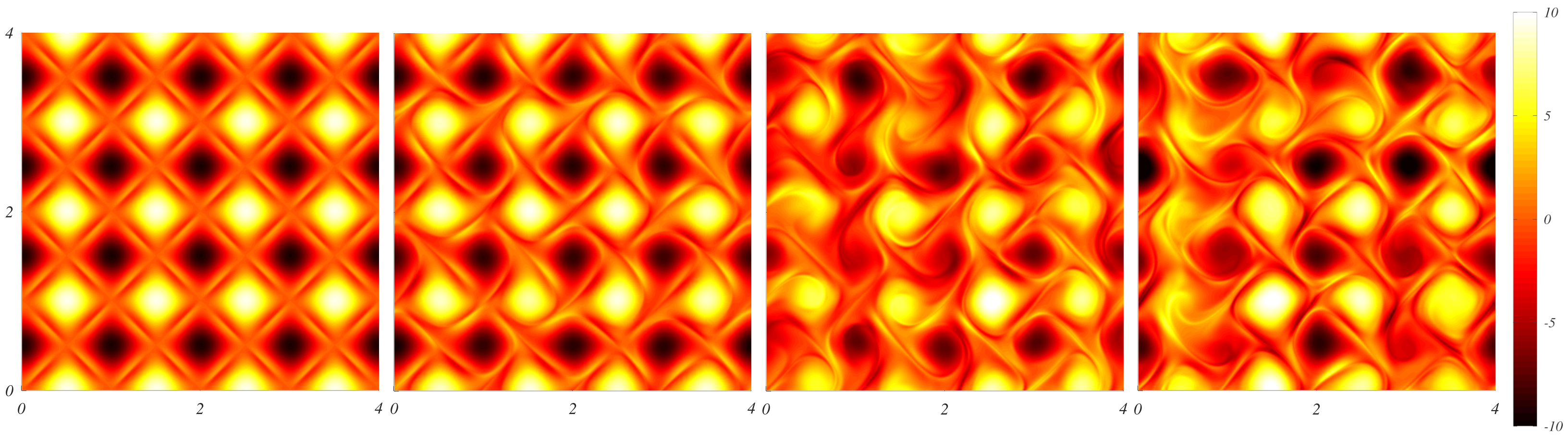}
\includegraphics[width=0.99\textwidth]{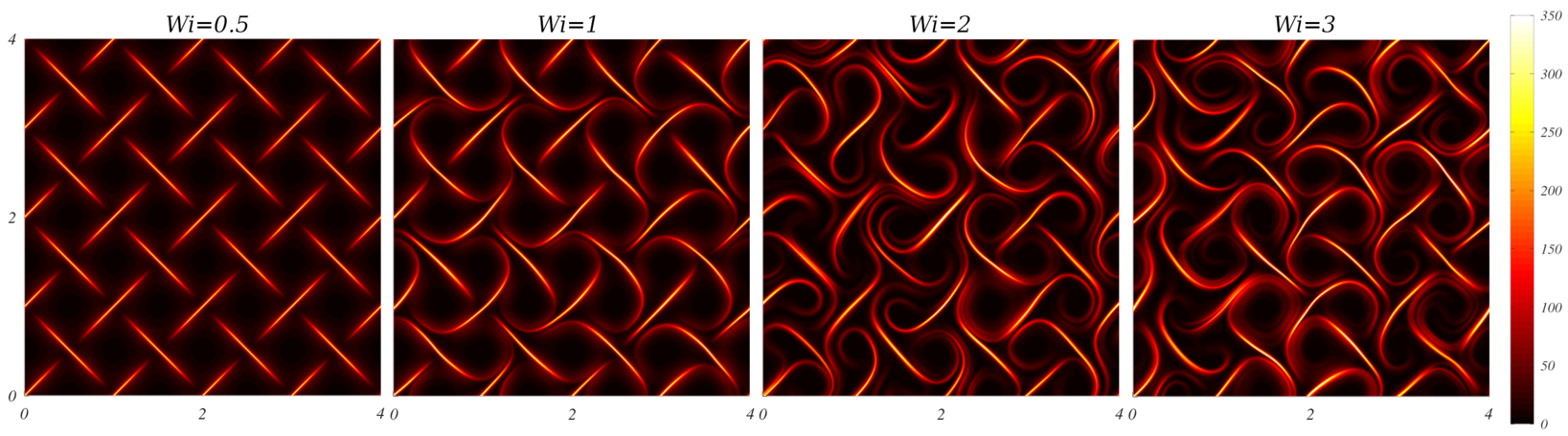}
\caption{Snapshots of the vorticity field (top), and normalised conformation tensor trace ($c_{ii}^{\left(1\right)}/Wi$) (bottom) for increasing $Wi$. Other parameters are ($Re=10^{-2}$, $\beta=1/2$, $\varepsilon=10^{-2}$, $\kappa=2.5\times{10}^{-5}$, $n=4$)\label{fig:snapshots_varwi}}
\end{figure}

\begin{figure}
\includegraphics[width=0.49\textwidth]{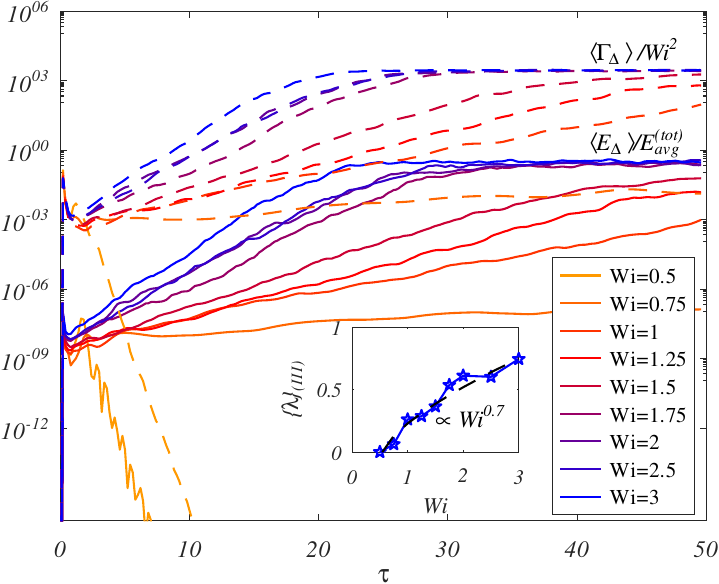}
\includegraphics[width=0.49\textwidth]{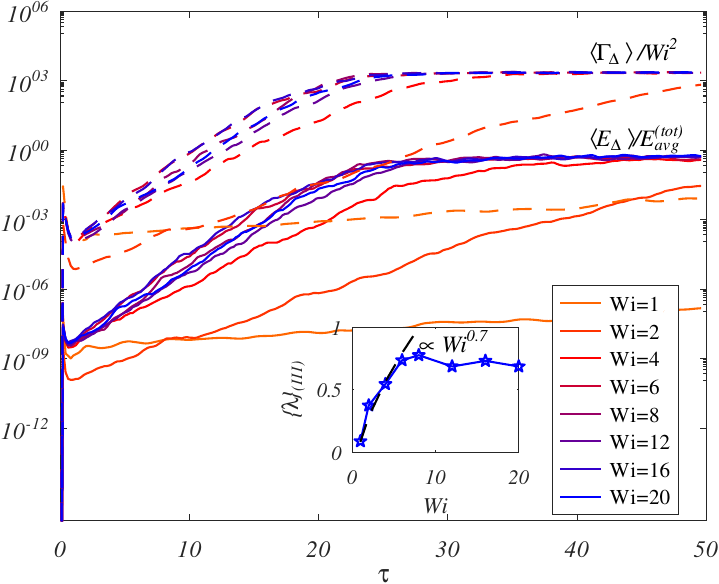}
\caption{Time evolution of $\left\langle{E}_{\Delta}\right\rangle/E_{avg}^{\left(tot\right)}$ and $\left\langle\Gamma_{\Delta}\right\rangle/Wi^{2}$, with increasing $Wi$. In the left panel, other parameters are ($Re=10^{-2}$, $\beta=1/2$, $\varepsilon=10^{-2}$, $\kappa=2.5\times{10}^{-5}$, $n=4$). In the right panel, the polymeric diffusivity is increased to $\kappa=10^{-4}$.\label{fig:dEG_varwi}}
\end{figure}

We next investigate the influence of elasticity on the evolution of uncertainty. Following the observation in~\S~\ref{sec:sc_nl} that for a given configuration, increased nonlinearity reduces the relative influence of polymeric diffusivity on the evolution of uncertainty, in this section, we take a base configuration with $\varepsilon=10^{-2}$ ($Re=10^{-2}$, $\beta=1/2$, $\varepsilon=10^{-2}$, $\kappa=2.5\times{10}^{-5}$, $n=4$), and vary $Wi$. Relative to the case with $\varepsilon=0$, we expect this to allow us to reach larger $Wi$ without the effects of polymeric diffusivity masking the nature of how the uncertainty evolution varies with $Wi$. Figure~\ref{fig:snapshots_varwi} shows snapshots of the vorticity field (upper panel) and normalised conformation tensor trace $c_{ii}^{\left(1\right)}/Wi$ (lower panel) for different values of $Wi$. For $Wi=0.5$, the flow is steady, the flow pattern is periodic on the scale of the forcing, and in a laminar regime. With increasing $Wi$ there is a symmetry breaking as the thin regions of high polymer deformation between cells interact with stagnation points and are swept laterally. Figure~\ref{fig:dEG_varwi} shows the evolution of $\left\langle{E}_{\Delta}\right\rangle/E_{avg}^{\left(tot\right)}$ and $\left\langle\Gamma_{\Delta}\right\rangle/Wi^{2}$ for a range of $Wi\in\left[0.5,3\right]$. \hla{In our simulations, at larger $Wi$ we obtain finer flow features, and hence t}o reach larger $Wi$, we require either a finer resolution, or larger values of polymeric diffusivity. The former will increase computational costs to a level beyond our resources for this work (doubling resolution increases costs by a factor of $>8$)\hla{ whilst for the latter option the chaotic dynamics will be altered by the polymeric diffusivity}. We show in the right panel of figure~\ref{fig:dEG_varwi} the evolution of $\left\langle{E}_{\Delta}\right\rangle/E_{avg}^{\left(tot\right)}$ and $\left\langle\Gamma_{\Delta}\right\rangle/Wi^{2}$ for a range of $Wi\in\left[1,20\right]$, with polymeric diffusivity increased to $\kappa=10^{-4}$. This increase in $\kappa$ permits stable simulations at larger $Wi$, but it is likely the dynamics of the flow are influenced by the polymeric diffusivity, particularly at larger elasticities. For smaller $Wi$ we see a sub-linear increase in the growth rate of uncertainty during the exponential regime (insets of both panels of figure~\ref{fig:dEG_varwi}). The dashed-black lines in both insets are linear in $Wi^{0.7}$, and for $\kappa=2.5\times10^{-5}$ in the left panel, there is a reasonable fit, suggesting the growth rate of uncertainty scales approximately with $Wi^{0.7}$. Acknowledging that Lyapunov dimension and Lyapunov exponent are different, we note that the Lyapunov dimension obtained by~\cite{plan_2017} was given in the form $CWi^{\alpha}$, with $\alpha\approx0.7$. At large $Wi$ (right panel of figure~\ref{fig:dEG_varwi}) this increase in $\left\{\lambda\right\}_{\left(III\right)}$ ceases, and for $\kappa=10^{-4}$ and $Wi\ge8$, the growth rate of uncertainty is roughly constant with further increases in $Wi$. We postulate that this limit is a consequence of the larger polymeric diffusivity, which acts to reduce uncertainty growth. For the simulations with larger $\kappa$, we observe a much more significant decrease of uncertainty at low $Wi$ during regime (II); in the right panel of figure~\ref{fig:dEG_varwi}, $\left\langle{E}_{\Delta}\right\rangle/E_{avg}^{\left(tot\right)}$ decreases significantly with decreasing $Wi$ during the early stages when polymeric diffusion dominates the uncertainty evolution. We note here that the rapid transfer of uncertainty across scales during regime (I) occurs for all $Wi$ investigated (even those at low $Wi$ in the laminar regime), and the growth rate or large scale uncertainty with $\tau^{6}$ in this regime holds across the range of $Wi$. 

\subsection{The effect of increasing domain size $n$}

Finally, we explore the influence of the domain size. We have observed in regime (I) rapid transfer of uncertainty across scales, and with the configuration used in this work, it is interesting to explore how uncertainty evolves differently as the domain size is changed. Note, that our non-dimensionalisation is based on the forcing, and not on any large scale flow features which may develop. An alternative non-dimensionalisation is possible, based on a fixed (unit) domain size, as was used in~\cite{plan_2017}. In that work they estimated the Lyapunov dimension for two different forcing wavelengths, and obtained a scaling with $Wi$ which was independent of the forcing scale. Denoting the dimensionless quantities based on the domain size with a subscript $n$, we can express the relationship $Re_{n}=nRe$, and $Wi_{n}=Wi/n$. If we non-dimensionalise based on the domain size, for fixed fluid transport properties (i.e. viscosity, relaxation times) and forcing magnitude, the Reynolds number would increase with domain size, and the Weissenberg number would scale with the inverse of domain size. Defining the Elasticity number as $El=Wi/Re$, we see the domain size based elasticity number $El_{n}$ scales with $1/n^{2}$. We conduct simulations all with the same (unit) forcing wavelength, for a range of domain sizes $n\in\left[2,16\right]$, and for three configurations: 1) with all other parameters matching the reference configuration ($Re=10^{-2}$, $\beta=1/2$, $\varepsilon=0$, $\kappa=2.5\times{10}^{-5}$, $Wi=2$); 2) using the sPTT model ($Re=10^{-2}$, $\beta=1/2$, $\varepsilon=10^{-2}$, $\kappa=2.5\times{10}^{-5}$, $Wi=2$); and 3) with the sPTT model and increased polymeric diffusivity ($Re=10^{-2}$, $\beta=1/2$, $\varepsilon=10^{-2}$, $\kappa=10^{-4}$, $Wi=2$). \hlc{Results highlighting the effect of sPTT nonlinearity parameter, $\varepsilon$ on uncertainty evolution are provided in Appendix~\mbox{\ref{sec:ptt}}.} Note that whilst the mean kinetic energy is independent of $n$, high-energy intermittent events increase with increasing $n$, and consequently for $n>10$, we use a smaller timestep of $\delta{t}=2\times{10}^{-4}$.

\begin{figure}
\includegraphics[width=0.49\textwidth]{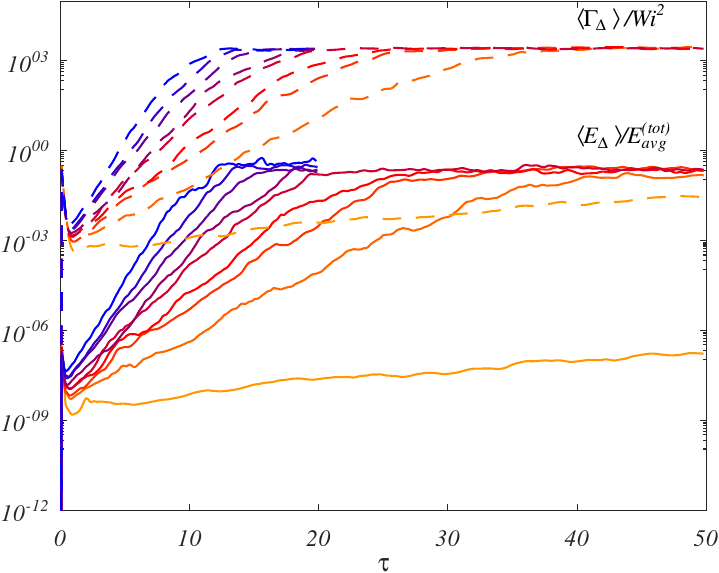}
\includegraphics[width=0.49\textwidth]{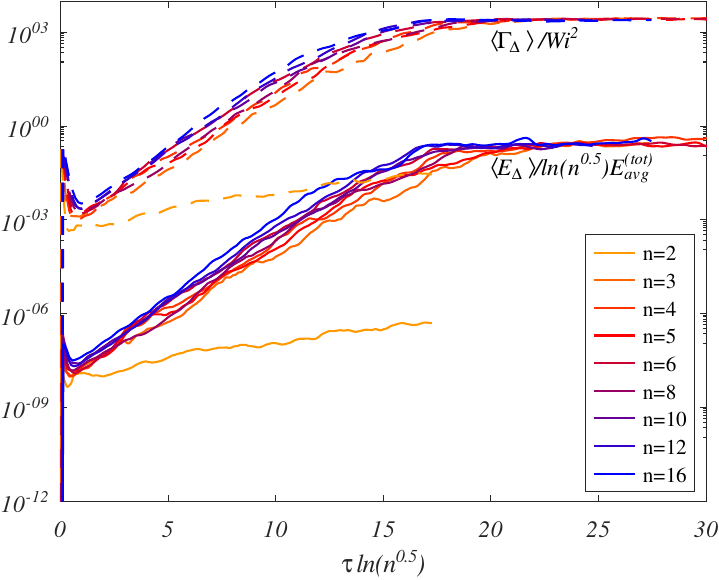}
\caption{Time evolution of $\left\langle{E}_{\Delta}\right\rangle/E_{avg}^{\left(tot\right)}$ and $\left\langle\Gamma_{\Delta}\right\rangle/Wi^{2}$ with increasing domain size $n$. Other parameters are ($Re=10^{-2}$, $\beta=1/2$, $\varepsilon=10^{-2}$, $\kappa=2.5\times{10}^{-5}$, $Wi=2$). On the left panel, $\tau$ has been rescaled by $\ln\left(n^{0.5}\right)$.\label{fig:dEG_varn}}
\end{figure}

The left panel figure~\ref{fig:dEG_varn} shows the evolution of $\left\langle{E}_{\Delta}\right\rangle/E_{avg}^{\left(tot\right)}$ and $\left\langle\Gamma_{\Delta}\right\rangle/Wi^{2}$ with increasing domain size $n$. All other parameters match configuration 2) described above ($Re=10^{-2}$, $\beta=1/2$, $\varepsilon=10^{-2}$, $\kappa=2.5\times{10}^{-5}$, $Wi=2$). In the right panel, the results are plotted against $\tau\ln{\left(n^{0.5}\right)}$. In the left panel, it is clear that increasing the maximum length scales of the flow results in a faster growth of uncertainty during regime (III). The collapse of the curves in the right panel of figure~\ref{fig:dEG_varn} shows this increased growth rate follows a trend of approximately $\ln\left(n^{0.5}\right)$. Our numerical experiments in which we increase $n$ for a fixed forcing strength may be re-framed as a set in which we increase $El_{n}^{-1/2}$ whilst co-varying $Re_{n}$ and $Wi_{n}$; we are effectively moving from (at small $n$) a high-$El_{n}$ regime where inertial effects are negligible, to (at large $n$) a low $El_{n}$ regime in which inertial effects play an increasing role. 

\begin{figure}
\includegraphics[width=0.49\textwidth]{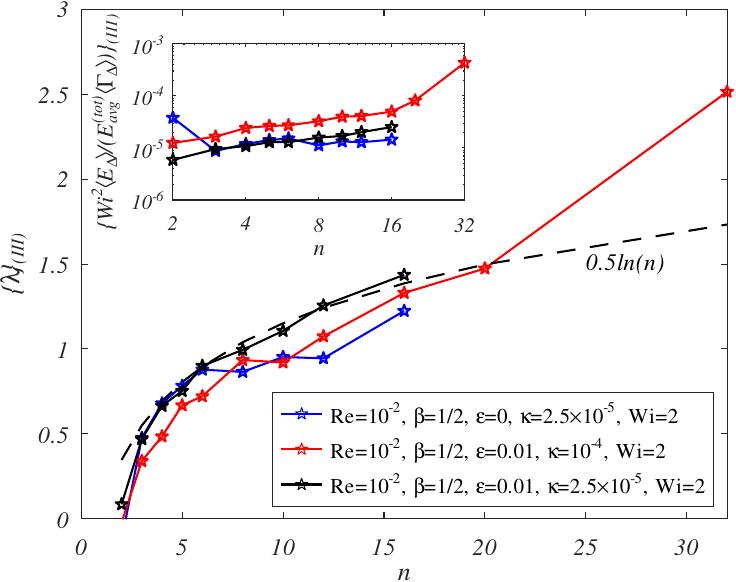}
\includegraphics[width=0.49\textwidth]{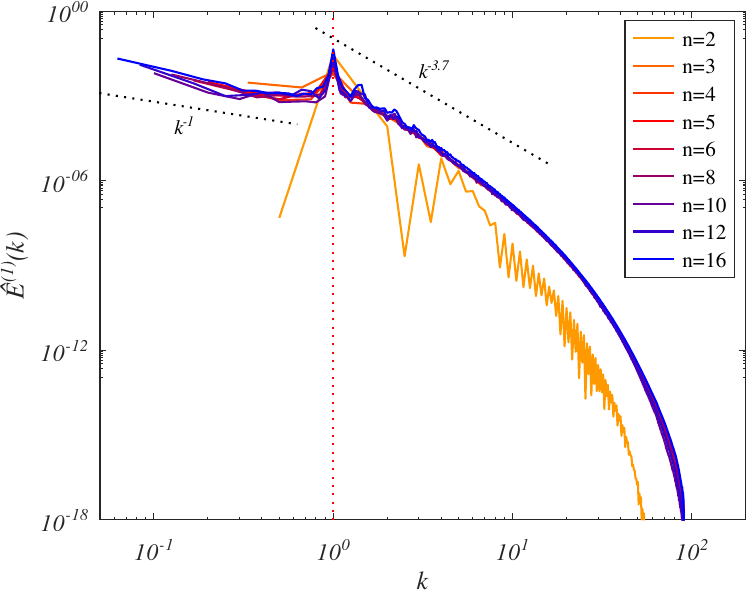}
\caption{Left panel: The variation of the growth rate $\left\{\lambda\right\}_{\left(III\right)}$ with increasing domain size $n$, for three different configurations. The dashed line corresponds to $0.5\ln{n}$. The inset shows the $\left\{Wi^{2}\left\langle{E}_{\Delta}\right\rangle/E_{avg}^{\left(tot\right)}\left\langle\Gamma_{\Delta}\right\rangle\right\}_{\left(III\right)}$. Right panel: the energy spectra of the reference field for different values of $n$. All other parameters match the reference configuration ($Re=10^{-2}$, $\beta=1/2$, $\varepsilon=0$, $\kappa=2.5\times{10}^{-5}$, $Wi=2$).\label{fig:growth_varn}}
\end{figure}

We denote the average growth rate during regime (III) as $\left\{\lambda\right\}_{\left(III\right)}$, and this is plotted against $n$ in figure~\ref{fig:growth_varn}, for several parameter configurations. For the cases with $\varepsilon=10^{-2}$ (red and black lines), we see that $\left\{\lambda\right\}_{\left(III\right)}$ scales with $0.5\ln\left(n\right)$ over the range $n\in\left[4,16\right]$. For small $n$, the growth rate drops below this. For the reference configuration (blue lines/symbols) the growth rate follows this trend up to $n=6$, and for larger $n$ drops below this trend. For the configuration with $\varepsilon=0.01$ and an increased $\kappa=10^{-4}$, a growth rate is lower (due to the effect of polymeric diffusivity to reduce uncertainty growth), but the variation with $n$ still follows the logarithmic trend. For the configuration with increased polymeric diffusivity ($Re=10^{-2}$, $\beta=1/2$, $\varepsilon=0.01$, $\kappa={10}^{-4}$, $Wi=2$), we can obtain converged results with a resolution of $\left(64n\right)^{2}$ modes, allowing us to reach larger domain sizes at reasonable computational costs. For $n=32$, the results deviate from the $0.5\ln\left(n\right)$ scaling. The inset of figure~\ref{fig:growth_varn} shows the variation of $\left\{Wi^{2}\left\langle{E}_{\Delta}\right\rangle/E_{avg}^{\left(tot\right)}\left\langle\Gamma_{\Delta}\right\rangle\right\}_{\left(III\right)}$ with $n$ for the three different configurations. In the Oldroyd B limit, this quantity represents the ratio of uncertainty in kinetic energy to that in elastic energy, and we see for all three configurations, this quantity increases with increasing domain size. This is despite the average energies of the reference field being independent of domain size.

\begin{figure}
\includegraphics[width=0.99\textwidth]{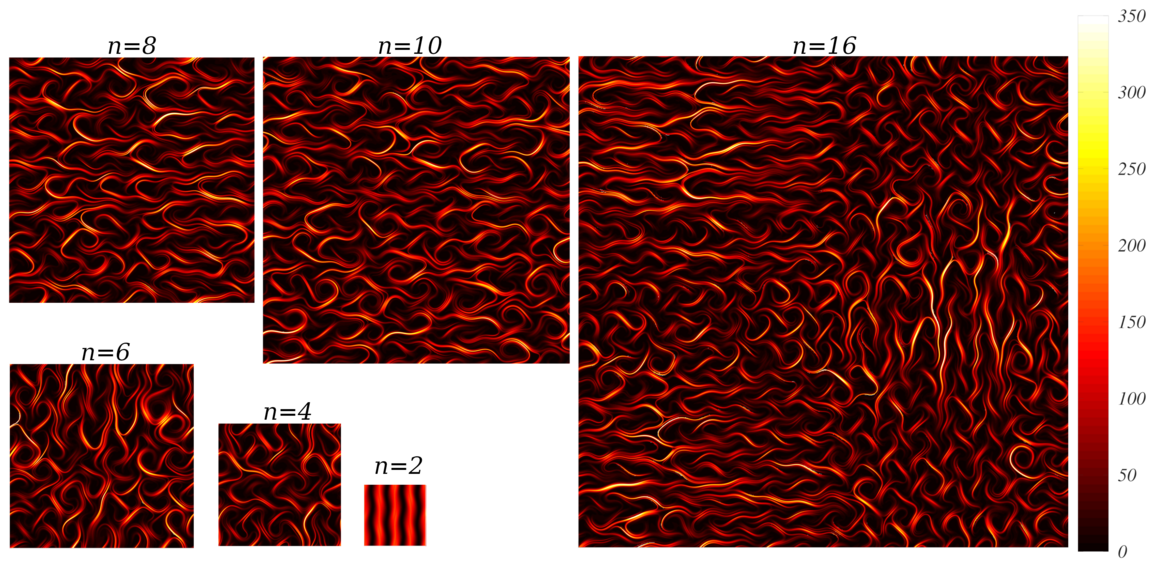}
\caption{Snapshots of the normalised conformation tensor trace ($c_{ii}^{\left(1\right)}/Wi$) for increasing $n$. All other parameters match the reference configuration ($Re=10^{-2}$, $\beta=1/2$, $\varepsilon=0$, $\kappa=2.5\times{10}^{-5}$, $Wi=2$).\label{fig:trc_varn}}
\end{figure}

It is known that in two-dimensional inertial turbulence, large scale condensates develop (see e.g.~\cite{chertkov_2007,svirsky_2023}) in finite domains, due to the inverse energy cascade and accumulation of energy at the largest scales. \hlc{In two-dimensional elasto-inertial turbulence, the inverse cascade is present~\mbox{\citep{gupta_2015}}, and such condensates may form in} settings with out-of-equilibrium forcing. The inverse cascade in two-dimensional (inertial, Newtonian) turbulence arises as a consequence of the additional constraint on the conservation of (squared) vorticity~\citep{svirsky_2024}. The addition of polymers provides a mechanism which can draw energy from large to small scales, weakening or reversing the cascade depending on the relative balance of inertial and elastic effects~\citep{gillissen_2019}. \hlc{In the present investigation, inertial effects are small, so the inverse cascade which leads to energy condensation is absent: the dynamics are a consequence of elastic effects drawing energy from large to small scales, with a scale-by-scale balance between this transfer and viscous dissipation.} The right panel of figure~\ref{fig:growth_varn} shows the energy spectra of the reference field for different domain sizes $n$. We see that there is energy below the forcing wavenumber, as in the simulations of~\cite{plan_2017}. In our simulations the energy spectra below the forcing wavenumber scales with $k^{-\alpha}$ for an $\alpha>1$; there must be \emph{some} transfer of energy from smaller to large scales. Furthermore, given the spectra above the forcing wavenumber closely match for all $n\ge3$, we postulate that it is the large scale flow structures which drive the increase in the rate of uncertainty growth with increasing domain size. Figure~\ref{fig:trc_varn} shows snapshots of the conformation tensor trace for increasing domain sizes. We see that for $n=2$, the flow is laminar. With increasing domain size, there are increasing regions exhibiting the arrowhead-like flow structures~\citep{page_2020}, and increasingly large scale patterns in the polymer deformation. A consequence of the rapid transfer of uncertainty across scales is that if one region of the flow has characteristics which result in a faster uncertainty growth, this will be \hla{transferred} throughout the domain increasing the uncertainty growth rate everywhere. In figure~\ref{fig:trc_varn} for $n=16$, we see some form of large scale structure consisting of alternating regions of arrowhead structures and cellular flow. \hlc{Whilst the mechanisms driving the formation of large scale flow features are uncertain, u}nderstanding the\hlc{ir} formation and structure may provide further insight into the dynamics of elastic turbulence\hlc{. S}uch investigations via numerical simulations become prohibitively expensive, and are left for future work. Note that when inspecting the conformation tensor trace fields for the case with sPTT nonlinearity $\varepsilon=10^{-2}$, we observe large scale patterns spanning the domain, but we do not observe the formation of arrowhead-like structures. 

\begin{figure}
\includegraphics[width=0.49\textwidth]{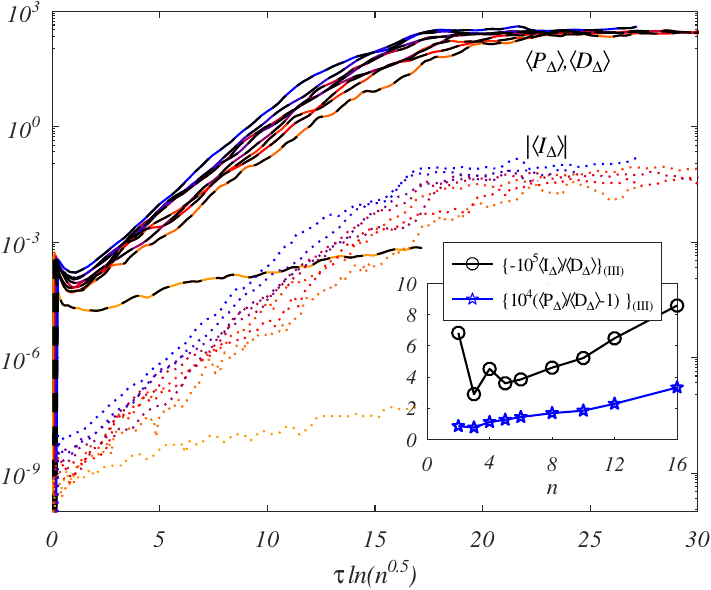}
\includegraphics[width=0.49\textwidth]{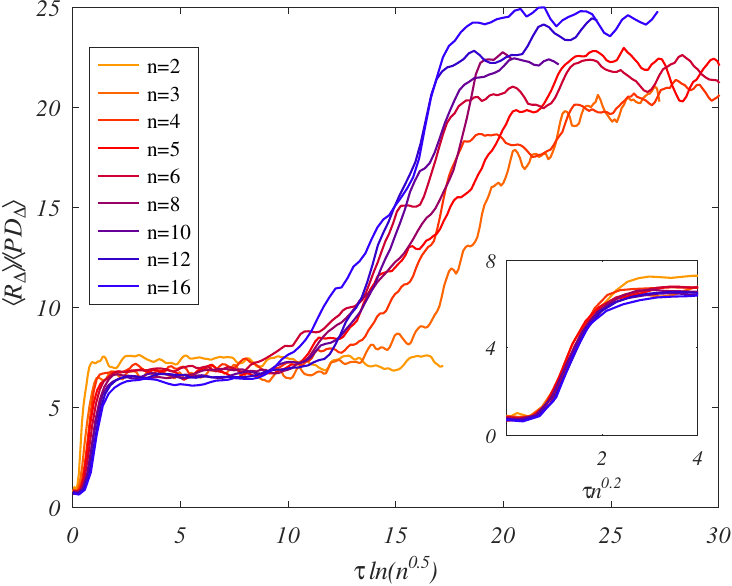}
\caption{Left panel: The time evolution of terms in~\eqref{eq:akeu} contributing to the evolution of $d\left\langle{E}_{\Delta}\right\rangle/dt$ for a range of domain sizes $n$. Time is scaled by $0.5\ln{n}$. The inset shows the variation of ratios of these terms with $n$. Right panel: The time evolution of the ratio $\left\langle{R}_{\Delta}\right\rangle/\left\langle{PD}_{\Delta}\right\rangle$, with time scaled by $0.5\ln{n}$. The inset shows the short time evolution of this ratio, which collapses with time scaled by $n^{0.2}$. Other parameters are ($Re=10^{-2}$, $\beta=1/2$, $\varepsilon=10^{-2}$, $\kappa=2.5\times{10}^{-5}$, $Wi=2$).\label{fig:dGdt_varn_2}}
\end{figure}

We next consider the terms in~\eqref{eq:akeu} and~\eqref{eq:dG} for varying $n$. The left panel of figure~\ref{fig:dGdt_varn_2} shows the individual terms in~\eqref{eq:akeu} for increasing $n$, plotted against $\tau\ln\left({n}^{0.5}\right)$. The same collapse is observed as in $\left\langle{E}_{\Delta}\right\rangle/E_{avg}^{\left(tot\right)}$ and $\left\langle\Gamma_{\Delta}\right\rangle/Wi^{2}$ in figure~\ref{fig:dEG_varn}. The inset shows the averages during regime (III) of ratios of these terms. Both quantities $\left\langle{I}_{\Delta}\right\rangle/\left\langle{D}_{\Delta}\right\rangle$, which represents the ratio of inertial production to viscous dissipation of uncertainty, and $\left\langle{P}_{\Delta}\right\rangle/\left\langle{D}_{\Delta}\right\rangle-1$, which indicates whether polymeric \hla{propagation} of uncertainty outweighs viscous dissipation, increase with increasing $n$, and this increase is roughly linear. However, the ratio $\left\langle{I}_{\Delta}\right\rangle/\left(\left\langle{P}_{\Delta}\right\rangle-\left\langle{D}_{\Delta}\right\rangle\right)$ decreases with increasing $n$, implying the relative influence of inertial effects on uncertainty growth decreases with increasing $n$. 

\begin{figure}
\includegraphics[width=0.49\textwidth]{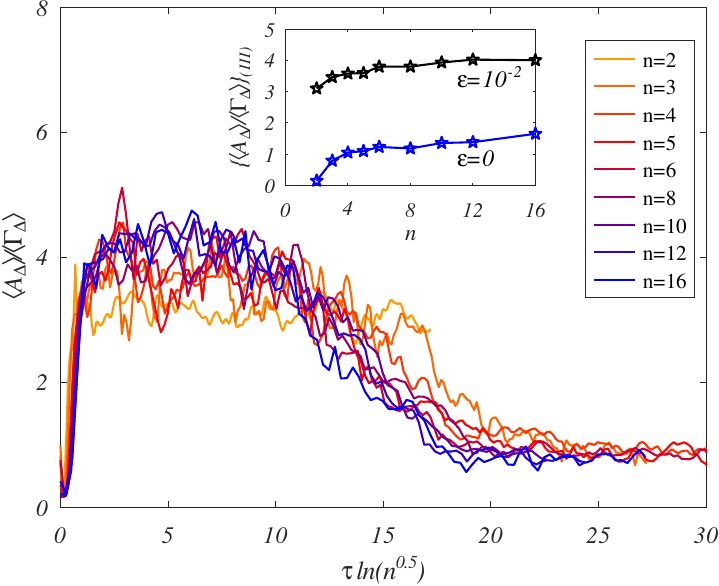}
\includegraphics[width=0.49\textwidth]{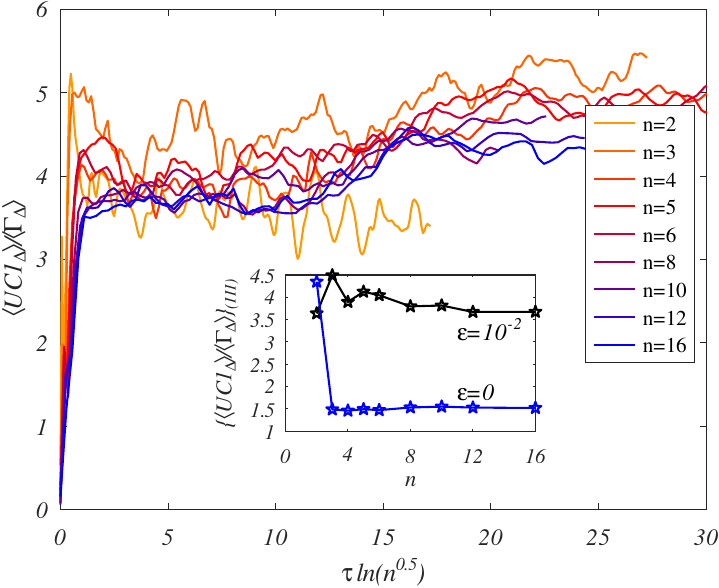}
\caption{The time evolution of terms in~\eqref{eq:dG} contributing to the evolution of $\left\langle\Gamma_{\Delta}\right\rangle$. The left panel shows the evolution of the ratio $\left\langle{A}_{\Delta}\right\rangle/\left\langle{R}_{\Delta}\right\rangle$. The right panel shows the evolution of the ratio $\left\langle{UC1}_{\Delta}\right\rangle/\left\langle{R}_{\Delta}\right\rangle$. In both cases, time is scaled by $0.5\ln{n}$. The insets show the average values of these ratios over the period of exponential growth. Other parameters are ($Re=10^{-2}$, $\beta=1/2$, $\varepsilon=10^{-2}$, $\kappa=2.5\times{10}^{-5}$, $Wi=2$).\label{fig:dGdt_varn}}
\end{figure}

The right panel of figure~\ref{fig:dGdt_varn_2} shows the evolution of the ratio $\left\langle{R}_{\Delta}\right\rangle/\left\langle{PD}_{\Delta}\right\rangle$, plotted against $\tau\ln\left(n^{0.5}\right)$. At short times, polymeric dissipation of uncertainty dominates in~\eqref{eq:dG}, and we see this in the inset with small values of $\left\langle{R}_{\Delta}\right\rangle/\left\langle{PD}_{\Delta}\right\rangle$ at early times. Note that we have plotted the data in the inset against $\tau{n}^{0.2}$ and the early time evolution of $\left\langle{R}_{\Delta}\right\rangle/\left\langle{PD}_{\Delta}\right\rangle$ collapses under this scaling. At the start of regime (III), there is an increase in $\left\langle{R}_{\Delta}\right\rangle/\left\langle{PD}_{\Delta}\right\rangle$ to approximately $6.5$, a value that is roughly independent of $n$, and persists through the exponential regime (III), before increasing further as a saturation of uncertainty is reached. In the left panel of figure~\ref{fig:dGdt_varn} we plot the evolution of $\left\langle{A}_{\Delta}\right\rangle/\left\langle{\Gamma}_{\Delta}\right\rangle$. The collapse with $\ln\left(n^{0.5}\right)$ still holds during regime (III), whilst (not shown), the collapse with $n^{0.2}$ observed for $\left\langle{R}_{\Delta}\right\rangle/\left\langle{PD}_{\Delta}\right\rangle$ holds. At late times, into regime (IV), for all $n$, $\left\langle{A}_{\Delta}\right\rangle/\left\langle{\Gamma}_{\Delta}\right\rangle$ tends towards the same value just below unity. During regime (III), the average value of $\left\langle{A}_{\Delta}\right\rangle/\left\langle{\Gamma}_{\Delta}\right\rangle$ (plotted in the inset) increases with increasing domain size. We see the same trend for the reference configuration with $\varepsilon=0$ (blue lines in the inset), but with lower magnitude. The right panel of figure~\ref{fig:dGdt_varn} shows the evolution of $\left\langle{UC1}_{\Delta}\right\rangle/\left\langle{\Gamma}_{\Delta}\right\rangle$. This ratio appears to be roughly independent of $n$ during regime (III). We know from~\eqref{eq:uc1} that $\left\langle{UC1}_{\Delta}\right\rangle$ depends on the relative orientation of the conformation tensor difference with the reference flow field, and the independence of $\left\langle{UC1}_{\Delta}\right\rangle/\left\langle{\Gamma}_{\Delta}\right\rangle$ with $n$ suggests changes in the domain size are not changing this orientation. We also note that for all three configurations for which we have conducted the investigation on increasing $n$, the magnitudes of both $\left\langle{R}_{\Delta}\right\rangle/\left\langle\Gamma_{\Delta}\right\rangle$ and $\left\langle{PD}_{\Delta}\right\rangle/\left\langle\Gamma_{\Delta}\right\rangle$ are independent of $n$ during regime (III). The changes we see in growth rate with increasing $n$ appear to be driven by increases in the production of uncertainty via polymeric advection, which couples uncertainty across scales, and corresponding increases in inertial and polymeric \hla{propagation} terms in the equation governing the evolution of $\left\langle{E}_{\Delta}\right\rangle$. 

\section{Conclusions}\label{conc}

%% Theory
In this work we have investigated the dynamics of uncertainty in elastic turbulence. Inspection of the evolution equations for uncertainty provides insight, showing a) that uncertainty in the polymeric deformation field can evolve into uncertainty in both flow and polymer fields, and hence for a chaotic flow exhibiting sensitivity to initial conditions, we would expect small perturbations or uncertainties inherent in polymer orientations to grow in finite time, with implications for the accuracy and repeatability of numerical and laboratory experiments. The \hla{evolution} of uncertainty depends on the relative alignments of the reference flow, the reference conformation tensor, the uncertainty in the flow and the uncertainty in the conformation tensor. The growth of uncertainty in kinetic energy is determined primarily by the balance of viscous dissipation and polymeric \hla{propagation}, whilst the growth of uncertainty in \hla{molecular deformation} is primarily controlled by the balance of production due to polymer advection, stretching and rotation, against polymer relaxation and polymeric diffusivity, with the latter two effects always acting to reduce uncertainty.

%% Regimes 
We identify several regimes of evolution of uncertainty.
\begin{enumerate}[(I)]
\item At very short times, a transfer of uncertainty across scales, with rapid growth of uncertainty in the flow and polymer deformation at large scales. Uncertainty at any point instantly generates uncertainty everywhere, as a consequence of the elliptic nature of the incompressibility constraint. Over a range of $Re$, this transfer of uncertainty from very small to very large scales approximately with $Re^{-2}$. In this regime uncertainty in the kinetic energy grows with time to the power of $6$ for large scales, and this growth rate appears to be independent of $Re$, $Wi$, polymeric diffusivity, the degree of non-linearity in the sPTT model, and the domain size.
\item At short times, of the order of one dimensionless time unit, a reduction in uncertainty, predominantly at small scales, but in some configurations across all scales, due to viscous and diffusive effects.
\item At moderate times, of the order of one to $30$ dimensionless time units, exponential growth of uncertainty across all scales. In the elastic turbulence regime, this growth rate decreases slightly with increasing Reynolds number, increases approximately with $Wi^{0.7}$, and increases with the logarithm of the square root of the maximum length scale. The growth rate decreases with increasing polymeric diffusivity, and is largely independent of the degree of nonlinearity in the sPTT model, within the range of nonlinearity which results in elastic turbulence. This regime is characterised by a slight rotation of the uncertainty in the polymeric deformation relative to the reference polymeric deformation, but the average rotation remains approximately constant throughout the regime, even as the magnitude of the uncertainty increases by six orders.
\item A late times, uncertainty saturates. In general, the maximum uncertainty in the kinetic energy, normalised by the total energy, is unity. For flows with a constant forcing, there is a limit to the extent to which the flows can decorrelate, and in the cellularly-forced elastic turbulence setting, this limit appears to be approximately $0.3$.
\end{enumerate}
At short times, in particular in regime (II), the uncertainty evolution is influenced by the nature of the initial uncertainty, but the growth rate in regime (III) is independent of the initial uncertainty: it is a function purely of the reference flow state. 

%% Comment on some implications.
The observation of rapid transfer of uncertainty across scales raises questions about closure models (analogous to large eddy simulation for inertial turbulence) for elastic turbulence. If uncertainty at the small scales can completely destabilise the trajectory of a flow across all scales, then any closure models which make assumptions about the flow at unresolved scales must somehow account for this. The observation also supports previous findings by~\cite{gupta_2019,yerasi_2024}, and highlights the need for care in numerical simulations of elastic turbulence, as numerical errors or inaccuracies may significantly alter the dynamics at large scales. 

%% Some future work type remarks
The approach taken in this study is new to viscoelastic flows, and there are many aspects we would like to investigate which are beyond the scope of the present paper. In particular, exploring the behaviour of uncertainty at extremely high elasticities, negligible polymeric diffusivities, and extremely large domain sizes would provide further insight, although such investigations require a finer resolution and will result in significantly increased computational costs. The formation of \hlc{large scale flow features} in settings where the maximum length scale is much larger than the forcing scale is a topic which has received little attention in the context of viscoelastic flows, but understanding this behaviour may have implications for the development of closure models for elastic turbulence. \hla{Our theoretical framework is limited to periodic flows, but it would be interesting to extend this approach to bounded flows (e.g. channel flows, or flows past obstacles). Whilst numerical experiments in such settings are straightforwards, developments to the theory are required to investigate the production and dissipation of uncertainty at walls.} We comment that the theory developed herein could equally be applied to flows with inertia, and a further study of three-dimensional elasto-inertial turbulence is planned. \hlb{Finally, the issue of polymeric dissipation in numerical simulations of chaotic viscoelastic flows is a topic of debate within the community, and although not the focus of the present study, we suggest that the present approach may provide useful insight on this issue - that diffusivity decreases uncertainty. Given that it is required for numerical stability, a detailed study on the effects of polymeric diffusivity on uncertainty evolution would be of value.}

\section*{Acknowledgements}
JK is funded by the Royal Society via a University Research Fellowship (URF\textbackslash R1\textbackslash 221290). We would like to acknowledge the assistance given by Research IT and the use of the Computational Shared Facility at the University of Manchester. We are grateful to three anonymous reviewers for their helpful suggestions from which our work has benefitted.

\section*{Declaration of Interests}
The authors report no conflict of interest.

\appendix

\section{The influence of resolution on uncertainty evolution}\label{sec:conv}

\begin{figure}
\includegraphics[width=0.49\textwidth]{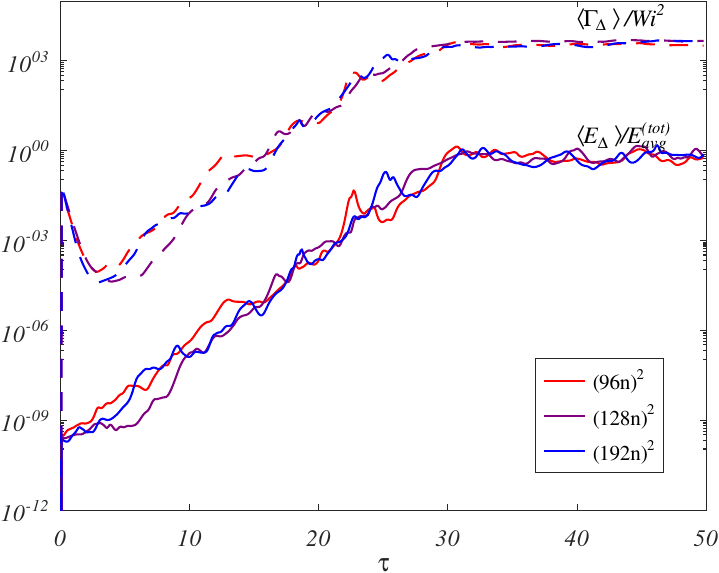}
\includegraphics[width=0.49\textwidth]{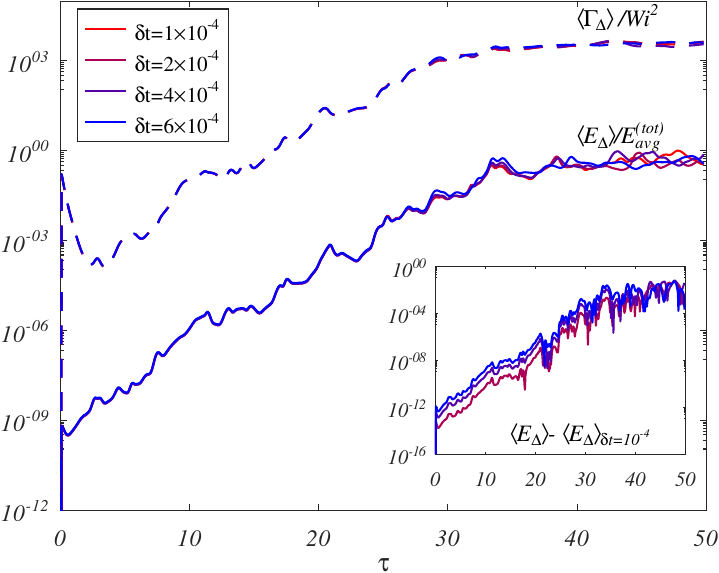}
\caption{\hla{The evolution of $\left\langle{E}_{\Delta}\right\rangle/E_{avg}^{\left(tot\right)}$ and $\left\langle\Gamma_{\Delta}\right\rangle/Wi^{2}$ for the reference configuration ($Re=10^{-2}$, $\beta=1/2$, $\varepsilon=0$, $\kappa=2.5\times{10}^{-5}$, $n=4$, $Wi=2$), for different spatial (left panel) and temporal (right panel) resolutions. In both panels, each line corresponds to an individual realisation, not an ensemble average. In the right panel, the inset shows the evolution of the difference between $\left\langle{E}_{\Delta}\right\rangle$, and $\left\langle{E}_{\Delta}\right\rangle$ for the smallest timestep $\delta{t}=10^{-4}$.}\label{fig:conv}}
\end{figure}

\hla{Beyond the comparison of the reference flow spectra for different resolutions provided in~\S~\mbox{\ref{ne}}, we here show the effects of temporal and spatial resolution on the evolution of uncertainty. The left panel of figure~\mbox{\ref{fig:conv}} shows the evolution of $\left\langle{E}_{\Delta}\right\rangle$ and $\left\langle\Gamma_{\Delta}\right\rangle$ for three different resolutions. Each realisation starts from a different precursor simulation (at a different resolution), but the evolution of $\left\langle{E}_{\Delta}\right\rangle$ and $\left\langle\Gamma_{\Delta}\right\rangle$ follow the same trend for all three resolutions. Importantly, the growth rate in regime (III) is the same for all three resolutions. The right panel of figure~\mbox{\ref{fig:conv}} shows the evolution of $\left\langle{E}_{\Delta}\right\rangle$ and $\left\langle\Gamma_{\Delta}\right\rangle$ for four different values of $\delta{t}$, with a resolution of $\left(128n\right)^{2}$ modes. Each realisation is initialised from the same precursor simulation, and the evolution of $\left\langle{E}_{\Delta}\right\rangle$ and $\left\langle\Gamma_{\Delta}\right\rangle$ match very closely for all values of $\delta{t}$ at short times. At late times (into regime (IV)), small discrepancies appear between the different timesteps. This is expected for a chaotic flow. In the inset of the right panel of figure~\mbox{\ref{fig:conv}}, we show the evolution of $\left\langle{E}_{\Delta}\right\rangle-\left\langle{E}_{\Delta}\right\rangle_{\delta{t}=10^{-4}}$, where $\left\langle{E}_{\Delta}\right\rangle_{\delta{t}=10^{-4}}$ is the uncertainty calculated with $\delta{t}=10^{-4}$; for all three values of $\delta{t}\in\left[2\times10^{-4},6\times{10}^{-4}\right]$, the evolution of this measure closely follows the evolution of uncertainty in the flow.}

\section{The effect of sPTT nonlinearity $\varepsilon$}\label{sec:ptt}

\begin{figure}
\includegraphics[width=0.49\textwidth]{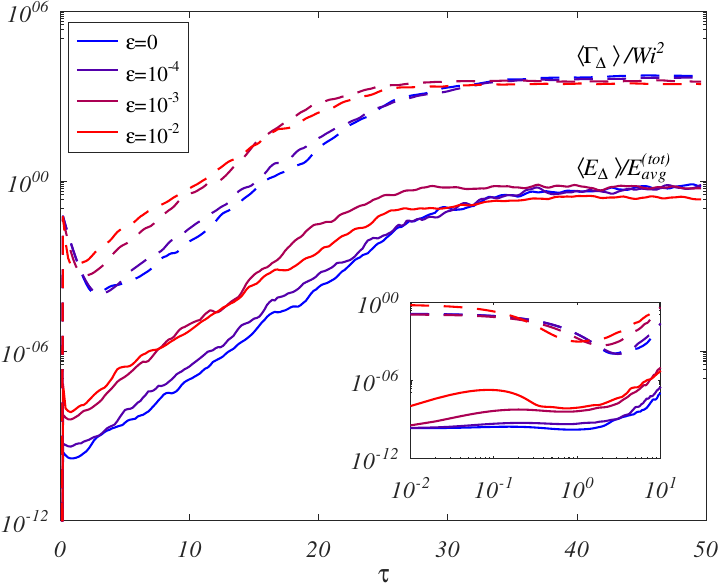}
\includegraphics[width=0.49\textwidth]{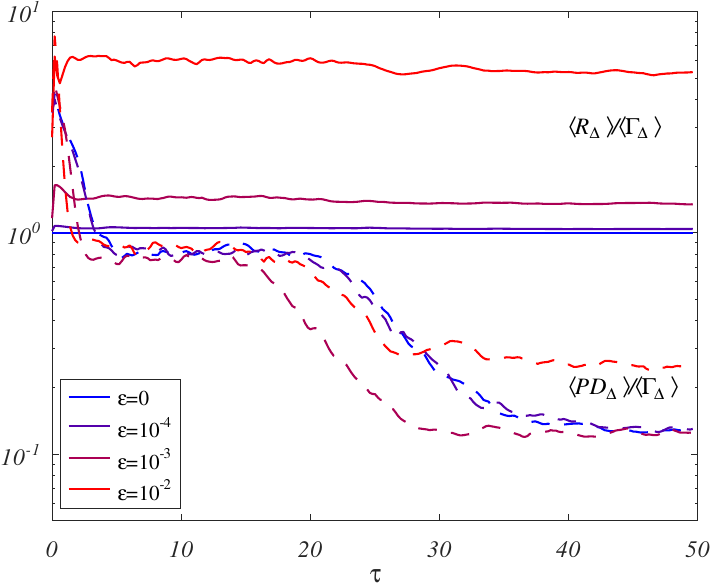}
\caption{Left panel: The evolution of $\left\langle{E}_{\Delta}\right\rangle/E_{avg}^{\left(tot\right)}$ and $\left\langle\Gamma_{\Delta}\right\rangle/Wi^{2}$ for a range of $\varepsilon$. Right panel: The evolution of ratios $\left\langle{R}_{\Delta}\right\rangle/\left\langle\Gamma_{\Delta}\right\rangle$ and $\left\langle{PD}_{\Delta}\right\rangle/\left\langle\Gamma_{\Delta}\right\rangle$. All other parameters match the reference configuration ($Re=10^{-2}$, $\beta=1/2$, $\kappa=2.5\times{10}^{-5}$, $n=4$, $Wi=2$).\label{fig:dEG_vareps}}
\end{figure}

The sPTT model exhibits shear thinning behaviour, and the extent of this shear thinning increases with increasing nonlinearity $\varepsilon$. The maximum extensional velocity is also influenced by the nonlinearity, scaling inversely with $\varepsilon$. To explore the influence of this nonlinearity on the uncertainty dynamics, we run simulations for $\varepsilon\in\left[0,10^{-4},10^{-3},10^{-2},10^{-1}\right]$, with all other parameters matching the reference configuration. For $\varepsilon=10^{-1}$ the flow remains laminar, and the mean kinetic energy is larger than for $\varepsilon=0$ by an order of magnitude. The left panel of figure~\ref{fig:dEG_vareps} shows the evolution of $\left\langle{E}_{\Delta}\right\rangle/E_{avg}^{\left(tot\right)}$ and $\left\langle\Gamma_{\Delta}\right\rangle/Wi^{2}$ for $\varepsilon\in\left[0,10^{-4},10^{-3},10^{-2}\right]$. A similar chaotic flow is observed for all $\varepsilon$ in this range, with an exponential growth rate in regime (III) approximately independent of $\varepsilon$. The saturation values of $\left\langle\Gamma_{\Delta}\right\rangle$ and $\left\langle{E}_{\Delta}\right\rangle$ in regime (IV) are changed to a small extent by the nonlinearity, with larger $\varepsilon$ slightly reducing these maximum values; increasing non-linearity reduces the maximum decorrelation achievable. We also observe some differences in the early time evolution of uncertainty (shown in the inset). For larger $\varepsilon$, the decrease in uncertainty in regime (II) is more pronounced, and occurs across all length scales in both the flow and polymer deformation fields (spectra calculated but not shown here for brevity). The right panel of figure~\ref{fig:dEG_vareps} shows the evolution of $\left\langle{R}_{\Delta}\right\rangle/\left\langle\Gamma_{\Delta}\right\rangle$ and $\left\langle{PD}_{\Delta}\right\rangle/\left\langle\Gamma_{\Delta}\right\rangle$. For $\varepsilon=0$, $\left\langle{R}_{\Delta}\right\rangle/\left\langle\Gamma_{\Delta}\right\rangle=1$ by definition. In all cases simulated, $\left\langle{R}_{\Delta}\right\rangle$ remains positive. With increasing nonlinearity, this ratio increases, and develops some temporal variation, although this temporal variation remains small. For $\varepsilon=10^{-2}$, $\left\langle{R}_{\Delta}\right\rangle/\left\langle\Gamma_{\Delta}\right\rangle\approx8$. During the exponential growth regime (III), the ratio $\left\langle{PD}_{\Delta}\right\rangle/\left\langle\Gamma_{\Delta}\right\rangle$ fluctuates, but remains in the range $\left[0.7,0.9\right]$ throughout. For all four values of $\varepsilon$, there is a decrease in $\left\langle{PD}_{\Delta}\right\rangle/\left\langle\Gamma_{\Delta}\right\rangle$ as regime (IV) is approached, and the final value of this ratio is larger (approximately $0.3$) for $\varepsilon=10^{-2}$, than for $\varepsilon\le10^{-3}$ (approximately $0.15$). We note that the ratio of $\left\langle{R}_{\Delta}\right\rangle/\left\langle{PD}_{\Delta}\right\rangle$ increases with increasing $\varepsilon$, suggesting that for a given configuration ($Re$, $Wi$, $\beta$, $\kappa$, $n$), increasing nonlinearity reduces the relative \hla{influence} of polymeric diffusivity on the evolution of uncertainty\hla{, although we add that this comment relates only to the sPTT model used here, and we do not make any inferences about other non-linear constitutive models (e.g. FENE-P). This observation suggests that for the sPTT model, increasing nonlinearity $\varepsilon$ may permit a larger value of polymeric diffusivity to be used without unduly influencing the chaotic dynamics of the flow}. Not shown, we note that $\left\langle{P}_{\Delta}\right\rangle/\left\langle{D}_{\Delta}\right\rangle-1$ is dependent on $\varepsilon$. For larger $\varepsilon$, the magnitude of the variation of $\left\langle{P}_{\Delta}\right\rangle/\left\langle{D}_{\Delta}\right\rangle-1$ in regimes (I) and (II) (shown in the inset of the right panel of figure~\ref{fig:dEdtref} for $\varepsilon=0$) increases, and the decrease in $\left\langle{E}_{\Delta}\right\rangle$ observed during regime (II) is more pronounced.

\begin{figure}
\includegraphics[width=0.99\textwidth]{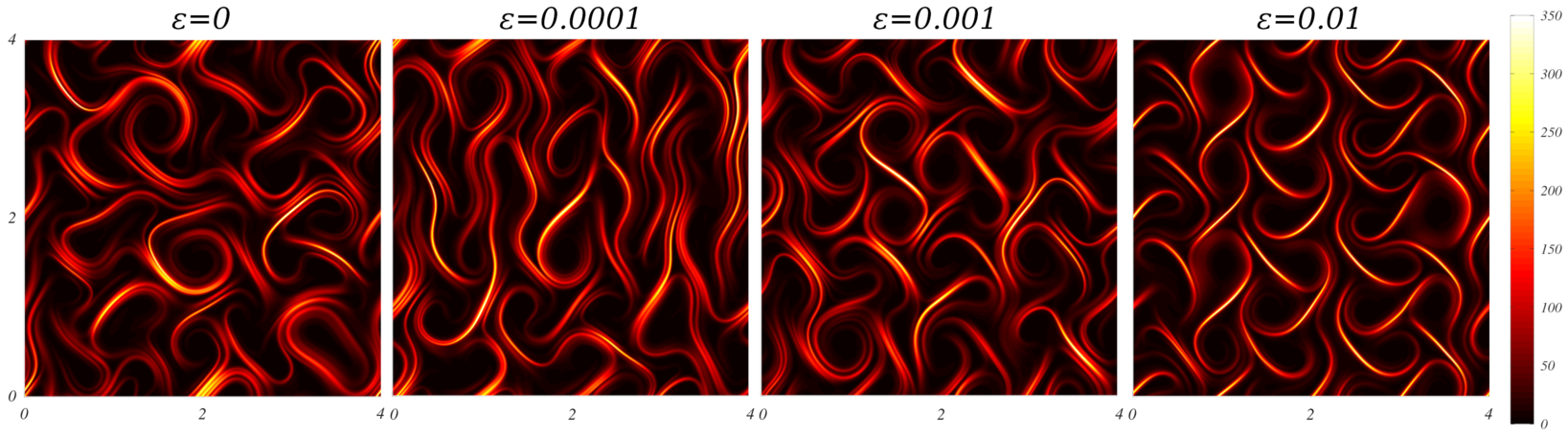}
\caption{Snapshots of the conformation tensor trace field for $\varepsilon\in\left[0,10^{-4},10^{-3},10^{-2}\right]$. All other parameters match the reference configuration ($Re=10^{-2}$, $\beta=1/2$, $\kappa=2.5\times{10}^{-5}$, $n=4$, $Wi=2$).\label{fig:trceps}}
\end{figure}

Figure~\ref{fig:trceps} shows snapshots of the conformation tensor field for a range of $\varepsilon$. For smaller $\varepsilon$ we see more large scale structures spanning the domain and breaking the cellular structure of the flow. For $\varepsilon=10^{-2}$, the cellular structure of the flow is much stronger. This is consistent with the larger value of $\left\langle{E}_{\Delta}\right\rangle/E_{avg}^{\left(tot\right)}$ obtained for smaller $\varepsilon$ in the saturation of uncertainty regime (IV). 

\section{An evolution equation for $\Pi_{\Delta}$}\label{sec:pi}

\hlb{As noted in~\S\mbox{\ref{ta}}, an alternative measure of the uncertainty in the polymer deformation can be given by the sum of the squares of the components of $\Delta{c}_{ij}$: $\Pi_{\Delta}=\Delta{c}_{ij}\Delta{c}_{ij}$. In two dimensions, this may be written $\Pi_{\Delta}=\Gamma_{\Delta}-2\det\left(\Delta{c}_{ij}\right)$. $\Pi_{\Delta}\ge0$ by construction, and is frame invariant. An evolution equation for $\Pi_{\Delta}$ may be derived as}
\begin{multline}
\frac{\partial\Pi_{\Delta}}{\partial{t}}+u_{k}^{\left(1\right)}\frac{\partial\Pi_{\Delta}}{\partial{x}_{k}}+\Delta{u}_{k}\frac{\partial\Pi}{\partial{x}_{k}}+2\Delta{c}_{ij}\Delta{u}_{k}\frac{\partial{c}_{ij}^{\left(1\right)}}{\partial{x}_{k}}
-2\Delta{c}_{ij}S^{\left(1\right)}_{ik}\Delta{c}_{kj}-2\Delta{c}_{ij}\Delta{S}_{ik}c_{kj}^{\left(1\right)}\\-2\Delta{c}_{ij}\Delta{S}_{ik}\Delta{c}_{kj}-2\Delta{c}_{ij}S_{jk}^{\left(1\right)}\Delta{c}_{ki}-2\Delta{c}_{ij}\Delta{S}_{jk}c_{ki}^{\left(1\right)}-2\Delta{c}_{ij}\Delta{S}_{jk}\Delta{c}_{ki}\\
-\frac{2\Pi_{\Delta}}{Wi}\left(1-d\varepsilon\right)+\frac{2}{Wi}\varepsilon\left(\Delta{c}_{kk}\right)^{2}-\frac{2\varepsilon}{Wi}\left(\Delta{c}_{ij}c_{ij}^{\left(1\right)}\Delta{c}_{kk}+\Pi_{\Delta}{c}_{kk}^{\left(1\right)}+\Pi_{\Delta}\Delta{c}_{kk}\right)\\
\kappa\frac{\partial^{2}\Pi_{\Delta}}{\partial{x}_{k}\partial{x}_{k}}-2\kappa\frac{\partial\Delta{c}_{ij}}{\partial{x}_{k}}\frac{\partial\Delta{c}_{ij}}{\partial{x}_{k}}
\end{multline}
\hlb{When we take an average over a spatially periodic domain, we obtain}
\begin{multline}
\frac{d\left\langle\Pi_{\Delta}\right\rangle}{dt}=\left\langle-2\Delta{c}_{ij}\Delta{u}_{k}\frac{\partial{c}_{ij}^{\left(1\right)}}{\partial{x}_{k}}\right\rangle+
\left\langle2\Delta{c}_{ij}S^{\left(1\right)}_{ik}\Delta{c}_{kj}+2\Delta{c}_{ij}\Delta{S}_{ik}c_{kj}^{\left(1\right)}+2\Delta{c}_{ij}\Delta{S}_{ik}\Delta{c}_{kj}\right\rangle\\+\left\langle2\Delta{c}_{ij}S_{jk}^{\left(1\right)}\Delta{c}_{ki}+2\Delta{c}_{ij}\Delta{S}_{jk}c_{ki}^{\left(1\right)}+2\Delta{c}_{ij}\Delta{S}_{jk}\Delta{c}_{ki}\right\rangle\\
-\left\langle\frac{2\Pi_{\Delta}}{Wi}\left(1-d\varepsilon\right)+\frac{2}{Wi}\varepsilon\left(\Delta{c}_{kk}\right)^{2}-\frac{2\varepsilon}{Wi}\left(\Delta{c}_{ij}c_{ij}^{\left(1\right)}\Delta{c}_{kk}+\Pi_{\Delta}{c}_{kk}^{\left(1\right)}+\Pi_{\Delta}\Delta{c}_{kk}\right)\right\rangle\\-\left\langle2\kappa\frac{\partial\Delta{c}_{ij}}{\partial{x}_{k}}\frac{\partial\Delta{c}_{ij}}{\partial{x}_{k}}\right\rangle,\label{eq:pi}
\end{multline}
\hlb{and note the similarities in construction of the advection term (first term on RHS), relaxation term (penultimate term) and diffusive term (final term) with the equivalent terms in~\mbox{\eqref{eq:dG}}. We note that an evolution equation for $\left\langle\det\left(\Delta{c}_{ij}\right)\right\rangle$ may be obtained by subtracting~\mbox{\eqref{eq:pi}} from~\mbox{\eqref{eq:dG}} and dividing by two.}

\section{Environmental impact of simulations}

Numerical simulations of the Navier-Stokes equations are computationally intensive, and have an associated carbon footprint. In our field this is rarely discussed or reported, but in the interests of sustainable research, a greater awareness of the environmental cost of simulations is essential. To this end, here we summarise the carbon footprint of simulations used in this study, with calculations performed using the Green Algorithms calculator~\citep{ga_2021}. All simulations were run on the Computational Shared Facility at the University of Manchester. For configurations with $n=4$ (at a resolution of $\left(128n\right)^{2}$ modes), a set of simulations (precursor and $10$ runs to obtain ensemble averages) required approximately $48$ hours on $64$ cores of an AMD EPYC Genoa 9634 CPU, with a carbon footprint of approximately $15.6$kgCO$_{2}$e. Larger simulations were run on up to $168$ cores, with the longest run time $360$ hours, and a carbon footprint of approximately $180.4$kgCO$_{2}$e. A small number of large simulations were run on Intel Xeon Gold 6130 CPUs, using up to $1024$ cores, with a maximum run time of $120$ hours, and the largest simulation having emissions of $454.7$kgCO$_{2}$e. The combined carbon footprint of \emph{all} simulations used in this work was approximately $9.1\times{10}^{3}$kgCO$_{2}$e. This estimate does not include the cost of post-processing and data analysis.

\bibliographystyle{jfm}
\bibliography{jrckbib}

\end{document}